\documentclass[aps,amsmath,amssymb,showpacs,showkeys]{revtex4-2}
\usepackage[dvips]{graphicx}
\usepackage{times}
\usepackage{braket}
\usepackage{xcolor}
\usepackage{adjustbox}
\usepackage{orcidlink}
\usepackage{hyperref}
\hypersetup{
  colorlinks=true,
  urlcolor=blue,
  linkcolor=red,
  citecolor=red
}
\begin{document}
\title{Magnetic Suppression of Cosmic Rays' Flux in $\boldsymbol{f(R)}$ and $\boldsymbol{f(Q)}$ Theories of Gravity}

\author{Swaraj Pratim Sarmah\orcidlink{0009-0003-2362-2080}}
\email[Email: ]{rs\_swarajpratimsarmah@dibru.ac.in}

\author{Umananda Dev Goswami\orcidlink{0000-0003-0012-7549}}
\email[Email: ]{umananda@dibru.ac.in}

\affiliation{Department of Physics, Dibrugarh University, Dibrugarh 786004, 
Assam, India}

%\date{}
\begin{abstract}
We investigate the effects of magnetic diffusion on the spectrum of ultra-high 
energy cosmic rays (UHECRs) from a cosmological perspective. To this end, we 
consider two modified theories of gravity (MTGs),  namely, the $f(R)$ gravity 
and a symmetric teleparallel gravity, also known as $f(Q)$ gravity. Utilizing 
these two MTGs, we calculate the suppression in the flux of UHECRs for a 
collection of sources. Non-evolution (NE) and cosmic star formation rate (SFR) 
scenarios have been considered in our calculation of the suppression factor. 
This study also includes a mixed composition scenario involving the nuclei 
upto iron (Fe). Furthermore, we provide a parameterization of the suppression 
factor for the proton and also for the mixed compositions within the $f(R)$ 
and $f(Q)$ theories, considering both NE and SFR scenarios. The influence of 
the turbulent magnetic field on the suppression factor is also incorporated 
in our work. Comparative analysis of all our results with the standard 
$\Lambda$CDM model reveals significant effects of MTGs on the suppression 
factor that the $f(R)$ power-law model predicts the lowest suppression factor,
while the $f(Q)$ model predicts the highest, and interestingly the results 
from the standard model fall within the range predicted by these two 
cosmological models.

\end{abstract}

%\pacs{}

\maketitle                                                                      

\section{Introduction}
In the annals of physics, the 1912 discovery of cosmic rays (CRs) by 
V.~F.~Hess stands as a pivotal chapter \cite{Hess}.  CRs are charged 
particles, consisting of protons, helium, and heavier ions up to iron. Despite 
more than a century-spanning quest since their unveiling, the origin and 
propagation mechanisms of CRs through the Universe largely remain shrouded in 
mystery \cite{harari,molerach,berezinskyGK}, with their narrative becoming most 
obscure beyond the energetic threshold of $E\geq 0.1$ EeV ($1$ EeV $= 10^{18}$ 
eV). The provenance of these enigmatic ultra-high energy CRs (UHECRs) eludes 
our grasp \cite{berezinskyGG, nagano, Bhattacharjee, Olinto}. Yet, for those 
within the more modest energy confines of $E < 0.1$ EeV, supernovae within 
the galactic bounds are believed to be the celestial forges of acceleration 
\cite{blasi201321, berezhko661, s.mollerach, Auger2013, TA2017}. Ascend higher, to realms around $1$ EeV 
and beyond, and one enters the extragalactic domain, are supposed to be 
accelerated by gamma-ray bursts or active galactic nuclei \cite{harari}.

The CR energy spectrum is a vast continuum, stretching from the GeV scale to 
the lofty heights of $100$ EeV, tracing a power-law spectrum. Along this 
spectrum, a subtle inflection known as the `knee' emerges at roughly 
$4 \times 10^{15}$ eV, and a `second knee' where the spectrum steepens more 
at around $10^{17}$ eV \cite{icecube2019, TA2020}, followed by a flattening at 
the `ankle' near $5 \times 10^{18}$ eV. And then, near $5 \times 10^{19}$ eV, 
one obtains the Greisen-Zatsepin-Kuzmin (GZK) cutoff—a stark demarcation in 
the CR spectrum \cite{Auger2021,icrc2021, Greisen, Zatsepin}. CRs are the 
silent voyagers of space, carrying energies that dwarf those produced by any 
human-made accelerator, a fact that has intrigued and challenged physicists 
\cite{harari_dark}. These particles, arriving at a rate of about one per square kilometer per century with energies exceeding $6 \times 10^{19}$ eV 
\cite{harari_dark}, are the messengers from the unknown, holding secrets to 
questions that have persisted for more than a century. The pursuit of understanding these cosmic wanderers has led to significant advancements in 
experimental physics, revealing a surprising decrease in the flux of CRs 
above $4 \times 10^{19}$ eV \cite{Auger2008}, contrary to earlier predictions. 
This suppression, confirmed by multiple studies \cite{harari_dark, hires2008, 
Auger2008, TA2013L1}, suggests a transition from lighter to heavier particles 
as energies increase, hinting at profound processes occurring over 
cosmological distances. It can also be understood by the source 
intensity population in the context of Hillas plot \cite{hillas,kp, mk}. This plot 
relates the maximum energy at which a source can accelerate particles to its size and 
magnetic field strength, highlighting that only certain astrophysical environments 
are capable of accelerating heavy nuclei to the observed energies \cite{kp}. Such 
observations, including the setting of the upper 
limits on the presence of photons \cite{harari_dark, Auger2009, TA2013}, 
neutrinos \cite{harari_dark, Auger2012L4, Icecube2013}, and neutrons 
\cite{harari_dark, Auger2012} among these UHECRs, have been crucial in our 
quest to understand the universe. The observations of a shift in composition 
at energies beyond $3 \times 10^{18}$ eV \cite{Auger2010, Auger2013JCAP} and a 
suppression in the flux above $4 \times 10^{19}$ eV, which could be indicative 
of the GZK cutoff—a theoretical prediction of energy loss over vast 
intergalactic travels, were made nearly sixty years ago 
\cite{Greisen, Zatsepin}. However, despite these insights, the complete 
story of CRs remains elusive, as current data is insufficient to confirm if 
energy loss during propagation is the sole cause of this suppression.

The vast intergalactic stage is set with turbulent magnetic fields (TMFs), 
playing a pivotal role in the propagation of UHECRs. These charged particles, 
propelled from distant galaxies, find their paths intricately influenced by 
the TMFs they encounter. The path of a UHECR through space is a delicate 
balance between its traversed distance and the scattering length 
$l_\text{D} = 3D/c$, where $D$ represents the diffusion coefficient and $c$ is 
the speed of light \cite{Supanitsky}. When the cosmic path is shorter than 
$l_\text{D}$, the particle's journey is ballistic. On the other hand, if the 
distance greatly exceeds $l_\text{D}$, the particle's movement becomes 
diffusive. The inclusion of extragalactic TMFs and the finite density of 
cosmic sources can lead to a magnetic horizon effect due to which the CR flux 
is suppressed for decreasing energies \cite{horizon_jcap}. This effect 
may reconcile observations with a higher spectral index 
\cite{s.mollerach, mollerach2013, wittkowski}, aligning with the theoretical 
values derived from diffusive shock acceleration. Another intriguing 
possibility involves the acceleration of heavy nuclei by extragalactic 
sources. These nuclei, upon interacting with the ambient infrared radiation 
(IR), undergo photodisintegration. This process releases a cascade of 
secondary nucleons, potentially explaining the lighter composition observed 
below the cosmic ankle \cite{unger, globus}. Intergalactic magnetic fields 
exert a profound influence on CRs, not only altering their trajectories but 
also potentially shaping their energy distribution. This latter effect, known 
as the magnetic horizon effect, manifests when magnetic fields are 
sufficiently intense to bar lower-energy CRs from reaching Earth, thereby 
reshaping the energy spectrum \cite{8, 9, 10, 11}. The extent of this 
influence hinges on the magnetic field's strength and the average spacing 
between CR sources. The propagation theorem \cite{aloisio} posits that 
magnetic fields do not impact the particle spectrum if source distances are 
significantly less than both the diffusion and attenuation lengths. 
Yet, in scenarios where magnetic fields are substantial and CR sources are 
sparsely scattered, a notable suppression in the low-energy end of the 
spectrum is anticipated. This suppression has previously been studied in 
Ref.~\cite{mollerach2013} at the low energy case. In Ref.~\cite{Manuel}, the 
authors have studied the suppression by both analytic and numerical 
approaches for the primary and secondary nuclei.

Albert Einstein's General Relativity (GR), formulated in 1915, stands as a 
pinnacle of theoretical physics, elegantly describing gravitational phenomena. 
A century later, the LIGO collaboration confirmed Einstein's prediction of 
Gravitational Waves (GWs) \cite{ligo}, and the Event Horizon Telescope captured
the first image of a black hole \cite{m87a, m87b, m87c, m87d, m87e, m87f}, 
both are monumental affirmations of GR. However, GR's inability to reconcile 
with quantum mechanics and account for the universe's accelerated expansion 
\cite{reiss, perlmutter, spergel, astier, Cognola:2007zu} with dark energy 
\citep{sami, udg_prd, Odintsov:2020nwm, Nojiri:2019fft, Odintsov:2019evb} and 
dark matter in galactic rotations \cite{Oort, Zwicky1, Zwicky2, Garrett, 
nashiba1} has led to the exploration of Modified Theories of Gravity (MTGs) 
\cite{Nojiri:2010wj, Nojiri:2017ncd}. Among these, the $f(R)$ gravity theory 
\cite{Sotiriou}, which replaces the Ricci scalar $R$ in the Einstein-Hilbert 
action by a function $f(R)$, has gained traction. Models like the Starobinsky 
\cite{starobinsky, staro}, Hu-Sawicki \cite{husawicki}, Tsujikawa 
\cite{tsujikawa}, and power-law \cite{powerlaw, udg_ijmpd} models of $f(R)$ 
gravity have been proposed to address these cosmic conundrums. Similarly, the 
symmetric teleparallel gravity and its extension, the $f(Q)$ gravity are 
modifications of the standard teleparallel gravity theory, where the 
non-metricity scalar $Q$ and its arbitrary function respectively replace the 
torsion scalar \cite{jimenez, harko, sanjay1, sanjay2, noemi}. The $f(Q)$ 
gravity has been used to investigate the cosmological implications, including 
the behaviour of dark energy, and has shown the potential to address 
cosmological tensions. Given MTGs' significant role in recent cosmological 
\cite{psarmah, gogoi_model} and astrophysical research 
\cite{jbora, nashiba2, ronit1, ronit_scripta, nashiba3, bidyut2024}, their 
application to UHECR studies, particularly in understanding the UHECR flux suppression, is a promising frontier. Till now, various research groups have
studied the anisotropy \cite{grapes3, globus2019, Yoshiguchi_2003, 
mollerach2022, mollerech2022b, Auger2017a, Abeysekara, merksch, m.ahlers, 
harari2021, erdogdu}, propagation mechanism \cite{mollerech2019, 
Sigl_1999, aloisio, bird, berezinsky_modif, berezinski_four_feat, berezinkyGre,
prosekin}, and suppression \cite{mollerach2013, Manuel, suppression1, 
suppression2} of UHECRs from the standard cosmology as well as the 
observatories \cite{hires, agasa, Auger2017, Auger2018, ta2019, augerprd2020, 
akeno_1993, hires2005, auger, auger2022, epjweb1, biermann_2012}. In our 
previous studies, we investigate the effects of $f(R)$ gravity on the UHECRs' 
propagation \cite{swaraj1} and anisotropy \cite{swaraj2} for a single source 
system. In this work, we extend our previous studies of UHECRs for the 
diffusive suppression of UHECR flux for an ensemble of sources and some 
scenarios of mixed composition of nuclei including primary, secondary, and 
their intermediate nuclei, focusing on the $f(R)$ and $f(Q)$ gravity models 
for the very first time and compares their implications on UHECRs' suppression 
with the result from standard $\Lambda$CDM model.

The structure of this paper is organized as follows: In Section \ref{secII}, 
we delve into the intricate process of CR diffusion in the context of a 
turbulent magnetic field. Section \ref{secIII} is neatly divided into two 
distinct parts. The first part introduces the $f(R)$ power-law model, while 
the other part presents a $f(Q)$ gravity model. In Section \ref{secIV}, we 
explore the suppression factor of the CR flux, leaving room for calculations 
involving multiple sources and a diverse composition of nuclei. The numerical 
results, along with their corresponding analytical fittings, are discussed in 
Section \ref{secV}. This is further divided into Subsection \ref{secVA} and 
Subsection \ref{secVB}, which deal with $f(R)$ and $f(Q)$ respectively. 
Finally, we wrap up our paper with a conclusive remark and a productive 
discussion in Section \ref{secVI}. 

\section{Diffusive Cosmic Rays in Turbulent Magnetic Fields}\label{secII}
Creating a model for extragalactic magnetic fields is difficult due to limited 
observations \cite{han}. The strength of these fields is not precisely 
known, which also varies across different regions \cite{hu_apj, urmilla}. 
Near the cluster centers, field strengths range from several to tens of 
\(\mu \text{G}\) \cite{han}. In less dense areas, they are weaker, typically 
between $1$ and $10$ nG, suggesting that significant fields may exist along 
the cosmic structures like filaments. Magnetic fields correlate with matter 
density, being stronger in dense regions such as superclusters and weaker in 
voids, possibly less than $10^{-15}$ G. Moreover, the magnetic fields
corelate with each other up to a maximum distance, known as the 
coherence length \(l_\text{c}\). Estimates suggest that in our Local 
Supercluster, magnetic fields have coherence lengths from $10$ kpc to $1$ Mpc 
and RMS strengths between $1$ and $100$ nG \cite{Sigl_1999}. The galactic 
magnetic field (GMF), with strengths of a few \(\mu \text{G}\), influences 
CRs' travel paths but has a limited effect on their overall spectrum 
due to its smaller scale. In our Local Supercluster, the presence of strong 
magnetic fields are indicated by the observed polarised rotations of background 
sources, with estimated strengths of 
$0.3$ to $2$ $\mu \text{G}$ \cite{molerach}. 
These fields are significant for CRs arriving from nearby sources. We simplify 
our study by focusing on CRs' movement through a uniform, 
turbulent extragalactic magnetic field. This field is characterized by its 
RMS strength \(B\) and coherence length \(l_\text{c}\), with \(B\) defined as 
\(\sqrt{\langle B^2(x)\rangle}\) and ranging from $1$ nG to $100$ nG 
\cite{feretti, Valle, Vazza}, while \(l_\text{c}\) spans from $0.01$ Mpc to 
$1$ Mpc \cite{sigl}. The magnetic field within the Local Supercluster is 
crucial for understanding CRs' arrival on Earth, as it is most influential 
for CRs from nearby sources. Therefore, we exclude the effects of larger 
structures like filaments and voids. 
%We simplify our study to CRs moving through a uniform and turbulent 
%extragalactic magnetic field, characterized by its RMS amplitude 
%$B_{\text{rms}}$ and coherence length $l_{\text{c}}$, which is the furthest 
%distance at which the field is correlated.

A critical energy $E_\text{c}$ can be defined as the energy where a charged 
particle's Larmor radius is equal to $l_{\text{c}}$, where the Larmor radius 
is given by 
%\begin{center}
$$r_\text{L}=E/Z|e|B \simeq 1.1 \frac{E/\text{EeV}}{Z B/ \text{nG}}\, \text{Mpc,}
$$
%\end{center}   
and hence the critical energy $E_{\text{c}}$ is given by 
$E_\text{c} = Z|e|B\, l_{\text{c}} = 0.9Z\left({B/\text{nG}}\right)\left({l_{\text{c}}/\text{Mpc}}\right) \text{EeV}$.
According to the MHD considerations $B(z)=(1+z)B(0)$ and the $l_c$ is 
assumed to be stretched by the expansion, so that $l_c(z) = l_c(0)/(1 + z)$ 
\cite{molerach}. Here, $B(0)$ and $l_c(0)$ are the present values of $B$ and $l_c$ 
respectively. $E_\text{c}$ distinguishes between two 
diffusion regimes: below $E_\text{c}$, resonant diffusion occurs, and above 
it, particles experience small deflections over $l_{\text{c}}$, leading to 
diffusion over much longer distances. When $ E < E_\text{c} $, the Larmor 
radius $ r_\text{L} $ is smaller than the coherence length $l_\text{c}$, 
leading to diffusion through resonant scattering at wavelengths comparable to 
\( r_\text{L} \). This is known as the quasi-linear regime. The diffusion 
length $l_\text{D}$ is influenced by the magnetic field's strength at the 
Larmor radius scale, which is dependent on the turbulent magnetic field 
spectrum. Generally, $ l_\text{D} \approx l_\text{c} \left(E/E_\text{c}\right)^\alpha $, where $\alpha $ varies with the spectrum: $ 1/3 $ for Kolmogorov, 
$ 1/2 $ for Kraichnan, and $1$ for Bohm diffusion, implying 
$l_\text{D} = r_\text{L} $. $l_\text{D} $ is the distance over which a 
particle typically deflects by $1$ radian, related to the random walk's basic 
step. The diffusion coefficient $ D(E)$ is $c\, l_\text{D}/3 $, and the mean 
square displacement $\langle (\Delta r)^2 \rangle $ after time $\Delta t $ is 
$6D\Delta t $ \cite{mollerach2013}.

For $ E > E_\text{c} $ ($ r_\text{L} > l_\text{c} $), scattering is 
non-resonant, caused by several small deflections with each 
$\sqrt{\delta \theta} \approx l_\text{c}/r_\text{L}$. After 
$N \approx l_\text{D}/l_\text{c}$ steps, the total deflection 
$ \Delta \theta \approx N \delta \theta = 1 $ radian, yielding 
$ l_\text{D} \approx l_\text{c} \left(E/E_\text{c}\right)^2$, indicating a 
rapid increase of diffusion length with energy. If $ l_\text{D} \ll r_\text{s}$
(the source distance), then spatial diffusion occurs; if 
$ l_\text{D} > r_\text{s}$, the particles propagate quasi-rectilinearly, 
causing minor angular diffusion. A perfect fit to the diffusion coefficient 
was given by Ref.~\cite{harari} as
\begin{equation}\label{d(e)}
D(E) \simeq \frac{c\,l_\text{c}}{3}\left[4 \left(\frac{E}{E_\text{c}} \right)^2 + a_\text{I} \left(\frac{E}{E_\text{c}} \right) + a_\text{L} \left(\frac{E}{E_\text{c}} \right)^{\alpha}   \right],
\end{equation}
and hence the diffusion length is given by
\begin{equation}
l_{D}(E) \simeq l_\text{c} \left[4 \left(\frac{E}{E_\text{c}} \right)^2 + a_\text{I} \left(\frac{E}{E_\text{c}} \right) + a_\text{L} \left(\frac{E}{E_\text{c}} \right)^{\alpha}   \right],
\end{equation}
where $a_\text{I}$ and $a_\text{L}$ are two constants. 
$a_\text{I} \approx 0.9$ and $a_\text{L} \approx 0.23$ according to the 
Kolmogorov spectrum, whereas as per the Kraichnan spectrum, 
$a_\text{I} \approx 0.65$ and $a_\text{L} \approx 0.42$ \cite{harari}.

The average separation between the sources of UHECRs, denoted as $d_\text{s}$, 
is intrinsically linked to their density $n_\text{s}$ and can be expressed as 
$d_\text{s} \approx n_\text{s}^{-1/3}$ \cite{mollerach2013}. For a value of 
$n_\text{s} = 10^{-3}~ \text{Mpc}^{-3}$, the separation is approximately 
$d_\text{s} \approx 10 ~ \text{Mpc}$. Consequently, for a lower density of 
$n_\text{s} = 10^{-6} ~ \text{Mpc}^{-3}$, the separation increases to 
$d_\text{s} \approx 100 ~ \text{Mpc}$. 

The shaping of CR spectra for different elements is fundamentally influenced 
by two main factors: their interactions with radiation backgrounds and the 
impact of cosmological evolution. Energy losses in CRs are due to adiabatic 
losses, represented as $dE/dt = -HE$, where $H = \dot{a}/{a}$ is the Hubble's 
parameter in terms of the scale factor $a$. These losses are prevalent across 
all energy levels and also result from interaction losses from various 
processes. Pair creation losses ($e^+ e^-$) become significant when 
interacting with Cosmic Microwave Background (CMB) photons of energy 
$\epsilon \sim 10^{-3}$ eV. This requires CR Lorentz factor $\Gamma > 10^9$, 
which implies that $E > A$ EeV for nuclei with mass number $A$. For protons, 
the energy loss length becomes less than the Hubble horizon 
$R_\text{H} = {c}/{H_0}$, when proton energies exceed $E_\text{p} > 2$ EeV. 
For heavier nuclei, due to an increased threshold and a larger pair production 
cross section ($\sigma \propto Z^2$), the energy loss length becomes 
comparable to $R_\text{H}$ when $E_\text{A} \approx A$ EeV \cite{allard}. The 
process of photo-pion production becomes noticeable only at extremely high 
energies, specifically when $E/A > 50$ EeV. Photo-disintegration of nuclei 
comes into play for $E > 2A$ EeV when interacting with CMB photons, and even 
at lower energies when interacting with higher energy background photons, such 
as those in the IR range. However, when interacting with IR photons, the 
energy loss length for photo-disintegration exceeds the Hubble horizon, except 
for heavier nuclei like Iron (Fe). For these heavier nuclei, the energy loss 
length is larger than $R_\text{H}$ only when $E < 30$ EeV \cite{allard}. 
Generally, for energies less than $E < Z$ EeV, the main energy losses are due 
to adiabatic effects, while interaction losses are minor \cite{mollerach2013}. Besides adiabatic 
losses, cosmological influences also contribute to the increase in CMB density 
and temperature with redshift. These influences may also affect the evolution 
of other radiation backgrounds, the density and emissivity of CR sources, and 
possibly the evolution of magnetic fields. Some additional mechanisms 
like the GZK effect, photodissociation, and pair production become 
increasingly important, especially for specific particle compositions and 
interactions with cosmic background radiation. The cosmological models 
also have some effects on the propagation of UHECRs. In the next section, we 
will discuss the cosmological models used for this work.
 
\section{Cosmological Models}\label{secIII}
In this section, we will discuss the cosmological models that will be used in 
the calculations for the different parameters needed for this current work. 
For this purpose, we consider one model from each of the two modified gravity theories: 
$f(R)$ and $f(Q)$. We will also introduce the Hubble parameter $H(z)$ in this 
section for the models of both modified gravity theories.

\subsection{$\mathbf{f(R)}$ gravity model}\label{secIIIA}
We consider here the simplest model, the $f(R)$ gravity power-law model, 
whose functional form is \cite{udg_ijmpd, d_gogoi, powerlaw}
\begin{equation}
f(R)= \beta R^n,
\end{equation}
where $\beta$ and $n$ are model parameters. The parameter $\beta$ has the 
dependence on the value of $n$ and other cosmological parameters such as 
Hubble constant $H_0$, current matter density parameter $\Omega_{\text{m}0}$ 
and also radiation density parameter $\Omega_{\text{r}0}$ via Ricci scalar 
at present time $R_0$ as given by \cite{d_gogoi}
\begin{equation}\label{lambda}
\beta = -\,\frac{3H_0^2\, \Omega_{\text{m}0}}{(n-2)R_0^n},
\end{equation} 
where the expression for $R_0$ is given by \cite{d_gogoi}
\begin{equation}\label{R0}
R_0 = -\, \frac{3 (3-n)^2 H_0^2\, \Omega_{\text{m}0}}{2n\left[(n-3)\Omega_{\text{m}0} + 2 (n-2) \Omega_{\text{r}0}\right]}.
\end{equation}

By using the Palatini formalism, the Friedmann equation in $f(R)$ theory in 
terms of redshift $z$ can be written as \cite{Santos}
\begin{equation}\label{fredmann}
\frac{H^2}{H_0^2}=\frac{3\,\Omega_{\text{m}0}(1+z)^3 + 6\,\Omega_{\text{r}0}(1+z)^4 + \frac{f(R)}{H_0^2}}{6 f'(R)\zeta^2 },
\end{equation}
where 
\begin{equation}\label{zeta}
\zeta = 1+ \frac{9f''(R)}{2f'(R)}\frac{H_0^2\, \Omega_{\text{m}0}(1+z)^3}{Rf'(R)-f'(R)}.
\end{equation}
Thus, from Eqs.~\eqref{fredmann} and \eqref{zeta}, we can express the Hubble 
parameter as a function of the redshift $z$ as
\begin{equation}\label{powerlawhubble}
H(z) = \left[-\,\frac{2nR_0}{3 (3-n)^2\, \Omega_{\text{m}0}} \Bigl\{(n-3)\Omega_{\text{m}0}(1+z)^{\frac{3}{n}} + 2 (n-2)\,\Omega_{\text{r}0} (1+z)^{\frac{n+3}{n}} \Bigl\}\right]^\frac{1}{2}.
\end{equation}
In our current study, we take the value of the model parameter $n=1.4$, which 
is the best-fitted value for the model as analyzed in Ref.~\cite{d_gogoi}. 
The values of the other cosmological parameters used in this study are 
taken as $H_0 \approx 67.4$ km s$^{-1}$ Mpc$^{-1}$ \cite{planck2018}, 
$\Omega_{\text{m}0} \approx 0.315$ \cite{planck2018}, and 
$\Omega_{\text{r}0} \approx 5.373 \times 10^{-5}$ \cite{nakamura}.
Thus the relation between cosmological time evolution and redshift can be 
expressed as \cite{swaraj1}
\begin{equation}\label{dtdz1}
\bigg | \frac{dt}{dz} \bigg |_{f(R)} =\frac{1}{(1+z)\, H} =(1+z)^{-1}  \left[-\,\frac{2nR_0}{3 (3-n)^2 \Omega_{\text{m}0}} \Bigl\{(n-3)\Omega_{\text{m}0}(1+z)^{\frac{3}{n}} + 2 (n-2)\Omega_{\text{r}0} (1+z)^{\frac{n+3}{n}} \Bigl\} \right]^{-\,\frac{1}{2}}\!\!.
\end{equation}
This relation will be used to calculate the CR flux for this model of $f(R)$
gravity in Section \ref{secIV}. 

\subsection{$\mathbf{f(Q)}$ gravity model}\label{secIIIB}
We consider the functional form of the $f(Q)$ gravity in this work 
as \cite{solanki} 
\begin{equation}
f(Q)= \sigma Q,
\end{equation}
where $\sigma$ is the model parameter. For this $f(Q)$ gravity model, the 
Hubble parameter in terms of the redshift $z$ can be written as \cite{solanki}
\begin{equation}\label{hzfq}
H(z) = H_0 \left[(1+z)^{\frac{3\sigma + C_1+C_2}{2 \sigma + C_2}} \left( 1+ \frac{C_0}{3\sigma + C_1+C_2} \right)- \frac{C_0}{3\sigma + C_1+C_2} \right],
\end{equation}
where $C_0$, $C_1$ and $C_2$ are three constant parameters, considered as bulk
viscous parameters. The best-fitted values of all these model parameters are 
$\sigma= -1.03^{+0.52}_{-0.55}$, $C_0=1.54^{+0.83}_{-0.79}$, 
$C_1=0.08^{+0.49}_{-0.49}$, and $C_2=0.66^{+0.82}_{-0.83}$ as reported in
the Ref.~\cite{solanki}. As in the previous case, the relation between 
cosmological time evolution and redshift can be expressed for this $f(Q)$ 
gravity model as
\begin{equation}\label{dtdz2}
\bigg | \frac{dt}{dz} \bigg |_{f(Q)} = (H_0(1+z))^{-1} \left[(1+z)^{\frac{3\sigma + C_1+C_2}{2 \sigma + C_2}} \left( 1+ \frac{C_0}{3\sigma + C_1+C_2} \right)- \frac{C_0}{3\sigma + C_1+C_2} \right]^{-1}.
\end{equation}
In Section \ref{secIV} we will use this relation to calculate the CR flux for 
this model of $f(Q)$ gravity.

\begin{figure}[!h]
\centerline{
\includegraphics[scale=0.5]{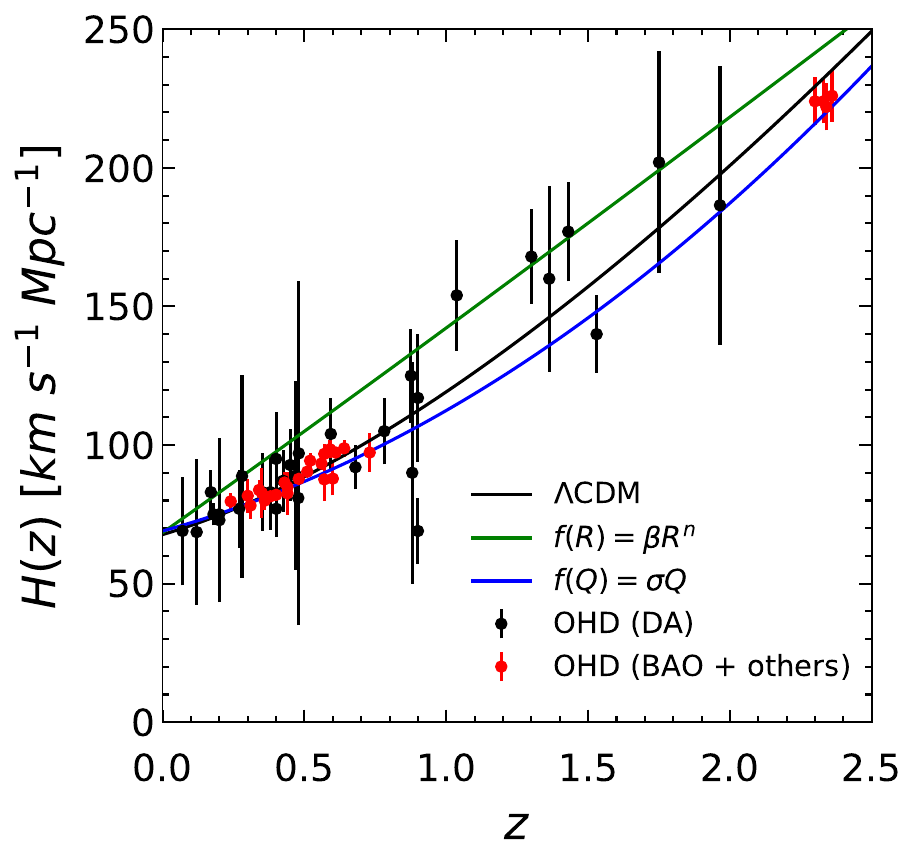}\hspace{1cm}
\includegraphics[scale=0.5]{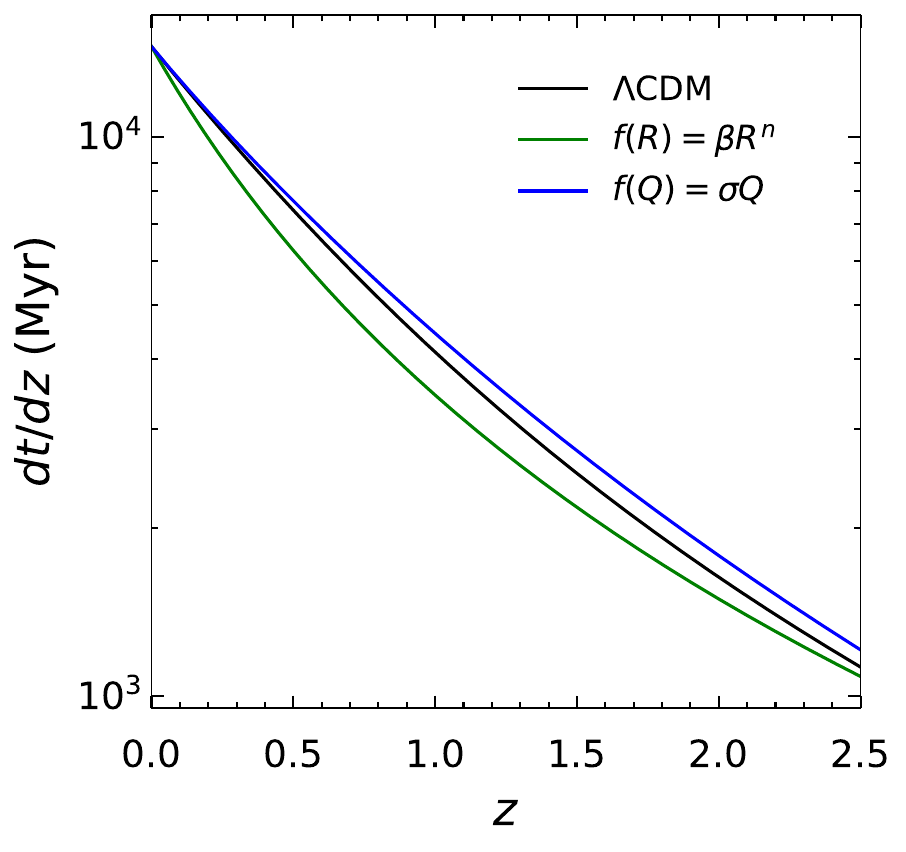}
}
\vspace{-0.3cm}
\caption{Left: Variations of Hubble parameter $H(z)$ with redshift $z$ as
predicted by the $\Lambda$CDM model, $f(R)$ gravity power-law model, and 
the $f(Q)$ gravity model in comparison with the observational Hubble data 
(OHD) obtained from differential age (DA) and Baryon Acoustic oscillations 
(BAO) methods \cite{solanki, swaraj1, d_gogoi}. Right: Evolution of cosmological time 
with $z$ as predicted by the aforementioned models of gravity.}
\label{fig0}
\end{figure}

In the left panel of Fig.~\ref{fig0}, we plot the Hubble parameters for the 
$f(R)$ gravity power-law model and the $f(Q)$ gravity model using 
Eqs.~\eqref{powerlawhubble} and \eqref{hzfq} respectively considering the 
respective set of model parameters as mentioned above along with the Hubble 
parameter for the standard $\Lambda$CDM model. We compare the results with 
the observational Hubble data (OHD) obtained from differential age (DA) and 
Baryon Acoustic oscillations (BAO) methods \cite{solanki, swaraj1, d_gogoi}. 
As one can see from this plot, both modified gravity models are well-fitted 
with the observational data. However, it needs to be mentioned that all three 
models predict slightly different values of Hubble constant $H_0$ as 
$67.77$ km s$^{-1}$ Mpc$^{-1}$ ($\Lambda$CDM), $68.4$ km s$^{-1}$ Mpc$^{-1}$ 
($f(R)$ power-law) and $69.0$ km s$^{-1}$ Mpc$^{-1}$ ($f(Q)$ model) 
respectively \cite{solanki}. It is clear that values for the $\Lambda$CDM and 
$f(R)$ power-law model are very close to the observed value of $H_0$ by the 
Planck experiment \cite{planck2018}. For completeness, we also plot the
cosmological time evolution with respect to redshift $z$ for all three models
of gravity considered as above in the right panel of Fig.~\ref{fig0}. From 
this plot, one can see that at the present value of redshift i.e. at $z=0$, 
there are no significant differences in the predictions of considered 
cosmological models. However, as the value of $z$ increases, the differences 
become more pronounced, especially with the $f(R)$ gravity power-law model. 
Whereas at $z=2.5$, the $\Lambda$CDM and $f(R)$ power-law models tend to 
converge, while the $f(Q)$ model follows a flat pattern.  

In the following section, we will apply these cosmological models to 
understand CR flux suppression. In this context, another pertinent point to 
be mentioned here is that variation of $H_0$ values for different models will 
not affect the CR flux suppression as it is independent of $H_0$. This will 
be clear in the following section.

\section{The Suppression of CR Flux}\label{secIV}
In the past, the diffusion of CRs in the TMF was discussed by several 
researchers \cite{berezinkyGre, blasi, globus, sigl, stanev, kotera, 
Yoshiguchi, lemonie1, hooper, hooper2, sigl2007, aloisio.ptep}. In the 
expanding Universe, a detailed analysis has been performed generalising 
the Syrovatskii solution \cite{syrovatsy_1959} for a proton diffusion by 
Berezinsky and Gazizov \cite{9, berezinkyGre}. The proton flux anticipated 
from a CR source at a distance of $r_\text{s}$, significantly exceeding the 
diffusion length $l_D$, can be determined by solving the diffusion equation 
within the expanding Universe \cite{berezinkyGre}. The resulting expression is 
as follows \cite{Manuel}:
\begin{equation}\label{flux}
J (E)=\frac{c}{4\pi}\, \int_{0}^{z_{\text{max}}}\!\! dz\, \biggl | \frac{dt}{dz}\biggl | \, \mathcal{L}\left[E_\text{g}(E, z), z\right] \frac{\textrm{exp}\left[-r_\text{s}^2/4 \lambda^2\right]}{(4\pi \lambda^2)^{3/2}}\, \frac{dE_\text{g}}{dE},
\end{equation} 
where $z_{\text{max}}$ is the maximum redshift of the source at which it starts
to emit CRs, $E_\text{g}(E, z)$ is the generation energy at redshift $z$ whose 
value is $E$ at $z=0$, $\mathcal{L}$ is the source emissivity that is obtained 
by summing up the contributions of the charge specific emissivity 
$\mathcal{L}_\text{Z}$ from various charges. We will employ a power-law with a 
rigidity cutoff $ZE_\text{max}$, represented as $\mathcal{L}_\text{Z}(E, z) = \xi_\text{Z} f(z) E^{-\gamma}/\cosh(E/ZE_\text{max}) $ \cite{mollerach2013}. Here
$\xi_Z$ represents the relative contribution to the CRs flux of charge $Z$ 
nuclei and $f(z)$ denotes the source emissivity evolution concerning 
the redshift $z$. Since 
H(z) varies across different cosmological models, the predictions made by these models will depend on the factor 
dt/dz, as described in Eqs. \eqref{dtdz1} and \eqref{dtdz2}. The Styrovatskii variable $\lambda^2$ is given by
\begin{equation}\label{syro}
\lambda^2(E,z)= \int_{0}^{z}dz\, \bigg | \frac{dt}{dz} \bigg |\,(1+z)^2 D(E_g,z).
\end{equation}     
Even though Eq.~\eqref{flux} was originally derived for protons, it can also 
be applied to nuclei by expressing it in terms of particle rigidities. When it 
comes to photo-disintegration processes in nuclei, they tend to conserve the 
Lorentz factor and rigidity of the main fragment. As a result, these processes 
do not significantly impact the particle's diffusion properties. However, 
there is a potential complication in relation to the photo-disintegration 
losses. As the source term $\mathcal{L}$ should refer to the primary nucleus 
that led to the observed one, obtaining this information is challenging 
due to the stochastic nature of the process. In the past, S.~Mollerach et 
al.~discussed this scenario \cite{mollerach2013} and we will extend this work 
in a modified gravity framework. Since in this work, we are interested in 
multiple sources instead of a single source, so according to the propagation 
theorem \cite{aloisio}, to sum all the sources, one can use 
\begin{equation}\label{lim1}
\int_{0}^{\infty} dr\, 4\pi r^2 \frac{\textrm{exp}\left[-r^2/4 \lambda^2\right]}{(4\pi \lambda^2)^{3/2}} =1. 
\end{equation}
To understand how the finite distance to the sources affects suppression, we 
compute the sum using a specific set of distance distributions. These 
distributions correspond to a uniform source density, and we consider the 
source distances from the observer as $r_\text{i} = (3/4\pi)^{1/3} d_\text{s} 
\Gamma (i+ 1/3)/(i-1)!$ \cite{mollerach2013, Manuel}, where $d_\text{s}$ is 
the distance between the sources and $i$ corresponds to the $i^{th}$ sources 
from a average distance. Hence, for a discrete source distribution, 
performing the sum over the sources, one can get a factor 
\cite{mollerach2013, Manuel} 
\begin{equation} \label{F_supp}
F \equiv \frac{1}{n_s} \sum_i \frac{\textrm{exp}\left[-r_\text{i}^2/4 \lambda^2\right]}{(4\pi \lambda^2)^{3/2}}
\end{equation}
instead of getting Eq.~\eqref{lim1}.

In Eq.~\eqref{flux}, after summing all the sources, we can write the combined 
flux for an ensemble of sources for the $f(R)$ gravity power-law model as
\begin{align}\label{flux_f(R)}
J_\text{com}(E) \Big |_{f(R)} \simeq  \frac{R_\text{H} n_\text{s}}{4\pi} \int_{0}^{z_\text{max}} dz\, (1+z)^{-1}  &\left[-\,\frac{2nR_0}{3 (3-n)^2 \Omega_{\text{m}0}} \Bigl\{(n-3)\Omega_{\text{m}0}(1+z)^{\frac{3}{n}} + 2 (n-2)\Omega_{\text{r}0} (1+z)^{\frac{n+3}{n}} \Bigl\} \right]^{-\,\frac{1}{2}} \nonumber \\[5pt]
&\times \mathcal{L}\left[E_\text{g}(E, z), z\right]\, \frac{dE_\text{g}}{dE}\, F,
\end{align}
where $R_\text{H} = c/H_0$ is the Hubble radius. Similarly, for the $f(Q)$ 
gravity model, the combined flux can be written as
\begin{align}\label{flux_f(Q)}
J_\text{com}(E) \Big |_{f(Q)} \simeq  \frac{R_\text{H} n_\text{s}}{4\pi} \int_{0}^{z_\text{max}} dz\,  (1+z)^{-1}  &\left[(1+z)^{\frac{3\sigma + C_1+C_2}{2 \sigma + C_2}} \left( 1+ \frac{C_0}{3\sigma + C_1+C_2} \right)- \frac{C_0}{3\sigma + C_1+C_2} \right]^{-1} \nonumber \\[5pt]
& \times \mathcal{L}\left[E_\text{g}(E, z), z\right]\, \frac{dE_\text{g}}{dE}\, F.
\end{align}
The final suppression factor of CR flux that describes the magnetic horizon 
effect can be expressed as
\begin{equation}\label{suppresion}
G(E/E_\text{c}) \equiv \frac{J_\text{com}(E)}{J_\text{com}(E) \big |_{d_\text{s} \rightarrow 0} },
\end{equation}
which is the ratio of the actual flux from discrete source distribution to that 
of the continuous source distribution ($d_\text{s} \rightarrow 0$). The 
continuous source distribution corresponds to the term $F = 1$ in 
Eqs.~\eqref{flux_f(R)} and \eqref{flux_f(Q)}, and it means that the flux is
independent of the modes of propagation of CRs.
Moreover, we can rewrite the Eq.~\eqref{syro} in terms of Hubble radius 
$R_\text{H}$ and from Eq.~\eqref{d(e)} as
\begin{equation}\label{ad}
\lambda^2(E,z)= H_0\frac{R_\text{H} l_\text{c}}{3}\int_{0}^{z}dz\, \bigg | \frac{dt}{dz} \bigg |\,(1+z)^2 \left[4 \left(\frac{(1+z)\,E}{E_\text{c}} \right)^2 + a_\text{I} \left(\frac{(1+z)\,E}{E_\text{c}} \right) + a_\text{L} \left(\frac{(1+z)\,E}{E_\text{c}} \right)^{\alpha}   \right].
\end{equation}
The suppression factor relies on the coherence length $l_\text{c}$ and the 
distance between the sources $d_\text{s}$ by this following relation \cite{mollerach2013, batista}
\begin{equation}\label{xs}
X_\text{s} = \frac{d_\text{s}}{\sqrt{R_\text{H} l_\text{c}}}.
\end{equation} 
This relation $X_\text{s}$ is known as the finite density factor and 
it will appear in the factor $F$ given in Eq.~\eqref{F_supp} after using
Eq.~\eqref{ad} in it. Hence $X_\text{s}$ will be used in Eq.~\eqref{flux_f(R)} 
and Eq.~\eqref{flux_f(Q)} while running the numerical calculations. In the 
next section, we will discuss the results obtained from the computations of 
Eq.~\eqref{suppresion} for different scenarios.

\section{Numerical Analysis}\label{secV}
This section is completely devoted to the numerical simulation and their 
corresponding analytic fitting. It is divided into two subsections 
for a systematic analysis: the first subsection for the $f(R)$ gravity model 
and the latter one for the model of $f(Q)$ gravity. We extensively use the 
\texttt{python scipy} library \cite{scipy} for the calculation with some 
assistance from the CRPropa 3.2 \cite{crpropa} and SimProp v2r4 
\cite{simprop} codes. Unless we specify, we use the RMS value of the 
magnetic field as $1$ nG, the coherence length of $1$ Mpc, and the spectral index 
$\gamma=2$ in all of the following plots. 

\subsection{Results from $\mathbf{f(R)}$ gravity model}\label{secVA}
\begin{figure}[!h]
\centerline{
\includegraphics[scale=0.42]{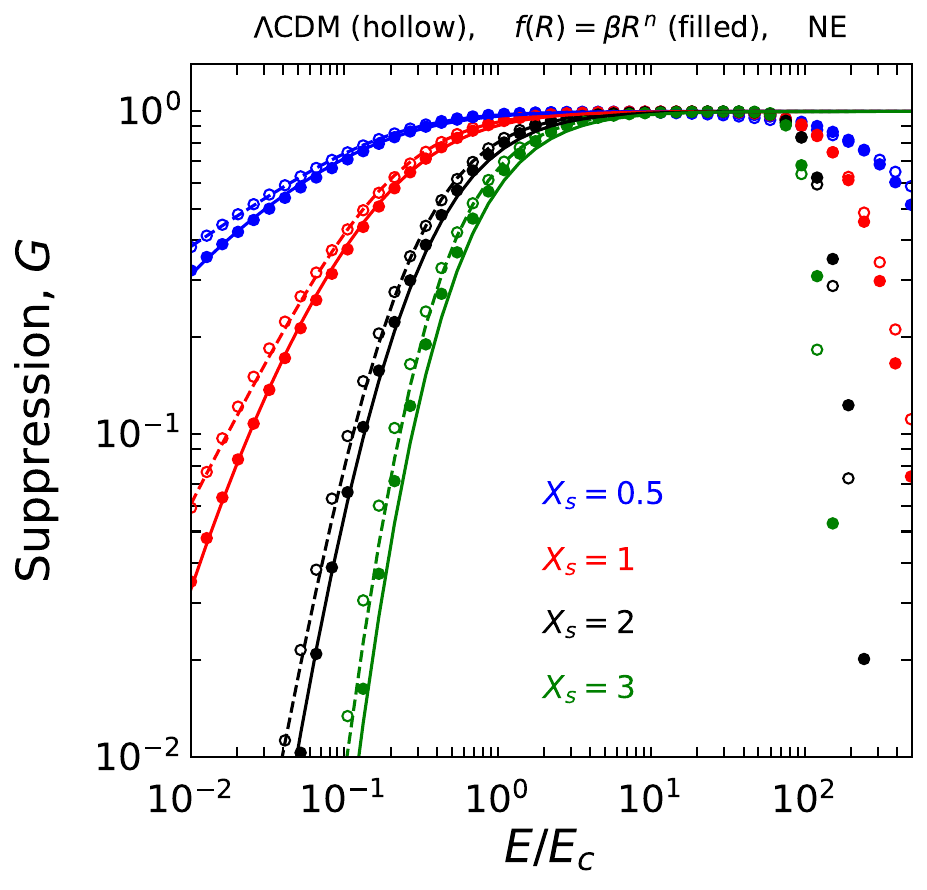}\hspace{1cm} 
\includegraphics[scale=0.42]{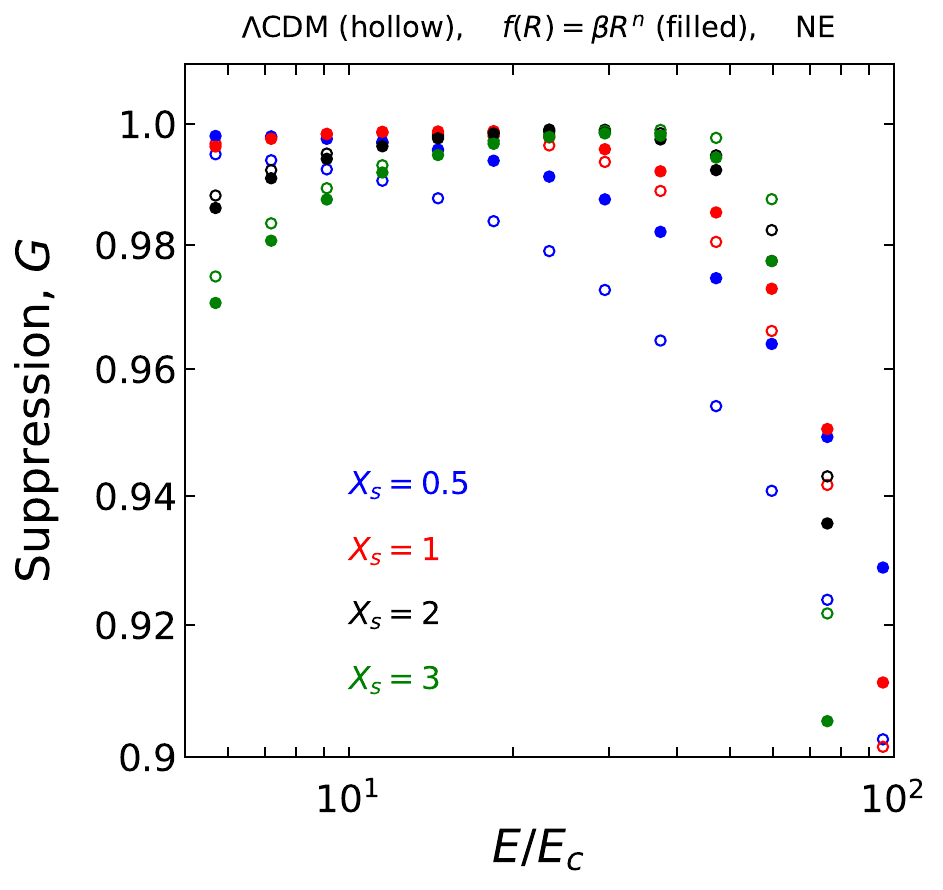}
}
\vspace{-0.3cm}
\caption{Left: Suppression factors for different finite density factors of 
sources $X_\text{s}$ as predicted by the $\Lambda$CDM model and the $f(R)$ 
gravity power-law model. The dotted and solid lines are for the analytic 
fitting for these two cosmological models respectively. Right: Same as the 
left panel but for a small scaling region. Both these plots are for the no 
cosmological evolution (NE) scenario. In the plots the hollow symbol 
represents the $\Lambda$CDM model and the filled symbol represents the $f(R)$
gravity power-law model. This representation will be followed in the rest of
similar figures.}
\label{fig1}
\end{figure}

In the left panel of Fig.~\ref{fig1}, we plot the suppression factor 
$G$ vs $E/E_\text{c}$ that is obtained by the magnetic horizon effect 
considering the maximum redshift of $z_\text{max}=2$. Here, we consider the 
finite density factor $X_\text{s} = 0.5, 1, 2$ and $3$. As this factor is 
increasing, the departure of the suppression factor is clearly seen. The 
filled dots represent the numerical simulation of the $f(R)$ model, while the 
hollow dots are for the $\Lambda$CDM model. The difference between the $f(R)$ 
model and the standard $\Lambda$CDM model is visible in this plot. The flux is 
less comparatively suppressed in the $f(R)$ gravity model. The solid lines 
represent the corresponding fits for the $f(R)$ gravity model, while the 
dotted lines for the $\Lambda$CDM model. We adopt the method of 
Ref.~\cite{Manuel} for the fitting formalism, but our parameterizations are a 
little bit different. The fitting equation that is used in the whole work is 
given by \cite{Manuel}          
\begin{equation}\label{fit}
P(x) = \text{exp}\left[-\left( \frac{a X_\text{s}}{x + b (x/a)^\eta}  \right)^\vartheta \right],
\end{equation}
where $a$, $b$, $\vartheta$ and $\eta$ are fitting parameters. The 
parameterizations required for fitting the expression throughout this work 
are given in Table \ref{table1}. In the right panel of Fig.~\ref{fig1},
we plot the same results as the left panel, but with a zoomed view in a 
small-scale region. It is seen that although the suppressions seem the same 
between the $E/E_\text{c}$ ranges of $1$ to $100$ in the left panel, in the 
zoomed view of the same on the right panel, they are not exactly the same.    
\begin{figure}[!h]
\centerline{
\includegraphics[scale=0.45]{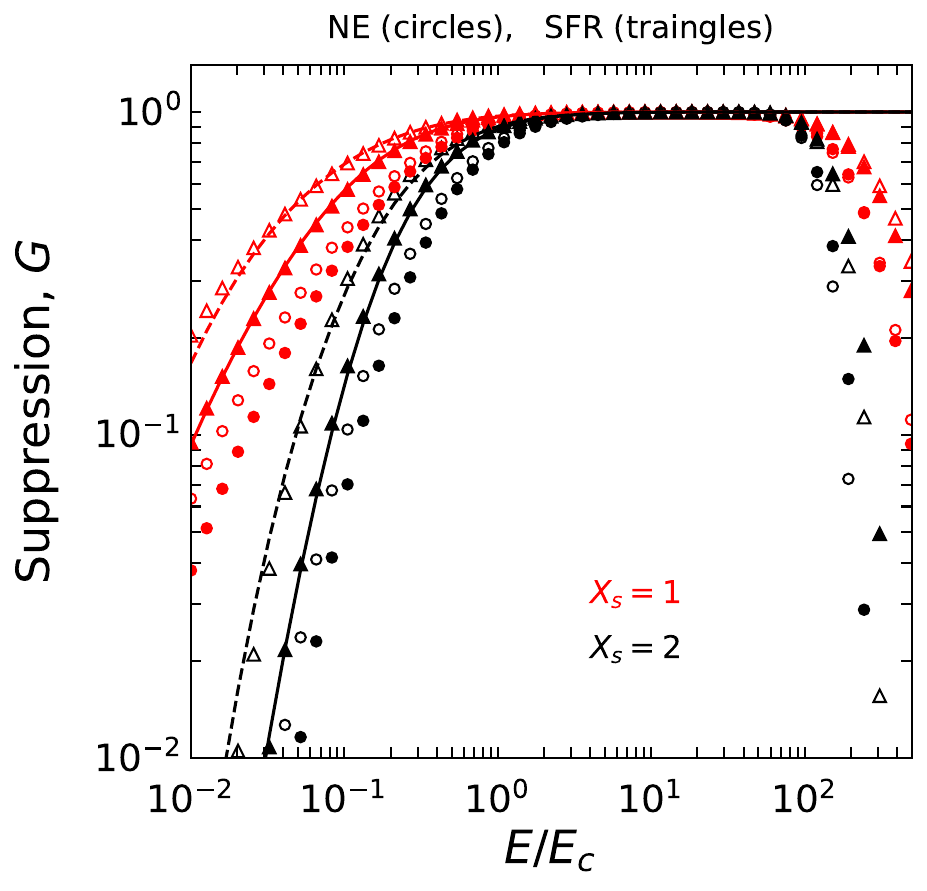} 
}
\vspace{-0.3cm}
\caption{Suppression factors for two different finite density factors of 
sources $X_\text{s}=1$ (red points) and $X_\text{s}=2$ (black points) as 
predicted by the $\Lambda$CDM model and the $f(R)$ gravity power-law model. 
The hollow and filled points represent the results from the $\Lambda$CDM model 
and the $f(R)$ gravity power-law model respectively. The 
circle and triangle in the plot represent the NE and SFR scenarios 
respectively. The solid and dotted lines represent the analytic fitting for 
the said cosmological models.}
\label{fig2}
\end{figure}

The suppression factor generally depends on the luminosity evolution of the 
sources with respect to the redshift. In Fig. \ref{fig1}, we have adopted the 
constant luminosity source evolution i.e.~no cosmological evolution (NE). 
For the cosmological evolution star formation rate (SFR) we adopt the 
formalism as follows \cite{hopkins}:
\begin{align}
\mathcal{L} \propto \left\{ \begin{array}{ll}
(1+z)^{3.44}, & \text{if  } z\leq 0.97 \\
(1+z)^{-0.26}, & \text{if  } 0.97 < z \leq 4.48 \\
(1+z)^{-7.8}, & \text{if  } z > 4.48
\end{array} \right.
\end{align}
For this SFR case, we simulate the suppression factor upto $z_\text{max} = 4$. 
In Fig.~\ref{fig2}, we plot the SFR scenario (filled and hollow triangle) 
along with the non-evolution (filled and hollow circle) case. Here, the hollow 
points represent the results from the $\Lambda$CDM model and those of filled 
ones represent the $f(R)$ gravity power-law model. For a certain energy, the 
cosmic particles arrive from a higher redshift in the SFR case than that of 
the NE case \cite{Manuel}. Here also the $f(R)$ gravity model is less 
suppressed in the lower energy region as well as in the higher energy range. 
The middle region suppression is similar (although not exactly the same as we 
have already shown in the right panel of Fig.~\ref{fig1}) for the both $f(R)$ 
model and the standard $\Lambda$CDM model. The solid and the dashed lines 
represent the fitting of Eq.~\eqref{fit} for the $f(R)$ and $\Lambda$CDM 
models respectively. The fitting parameters for the $f(R)$ power-law model and 
$\Lambda$CDM model each with the NE and SFR cases are given in Table 
\ref{table1}.

\begin{figure}[htb!]
\centerline{
\includegraphics[scale=0.42]{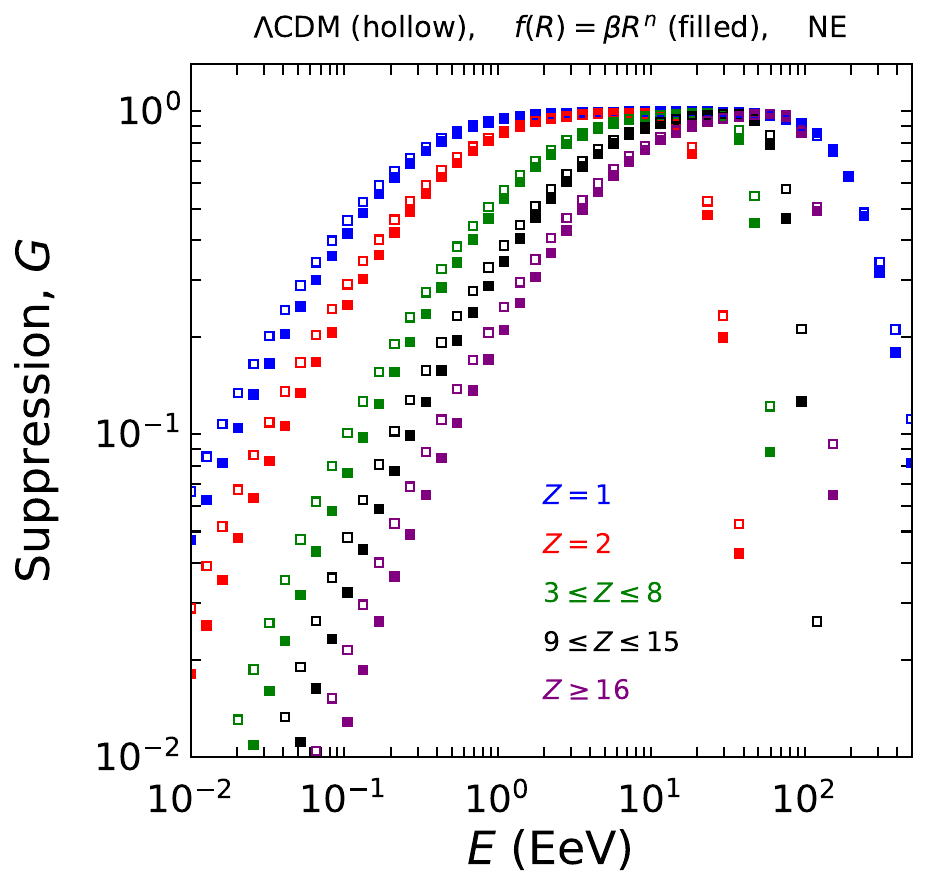}\hspace{1cm} 
\includegraphics[scale=0.42]{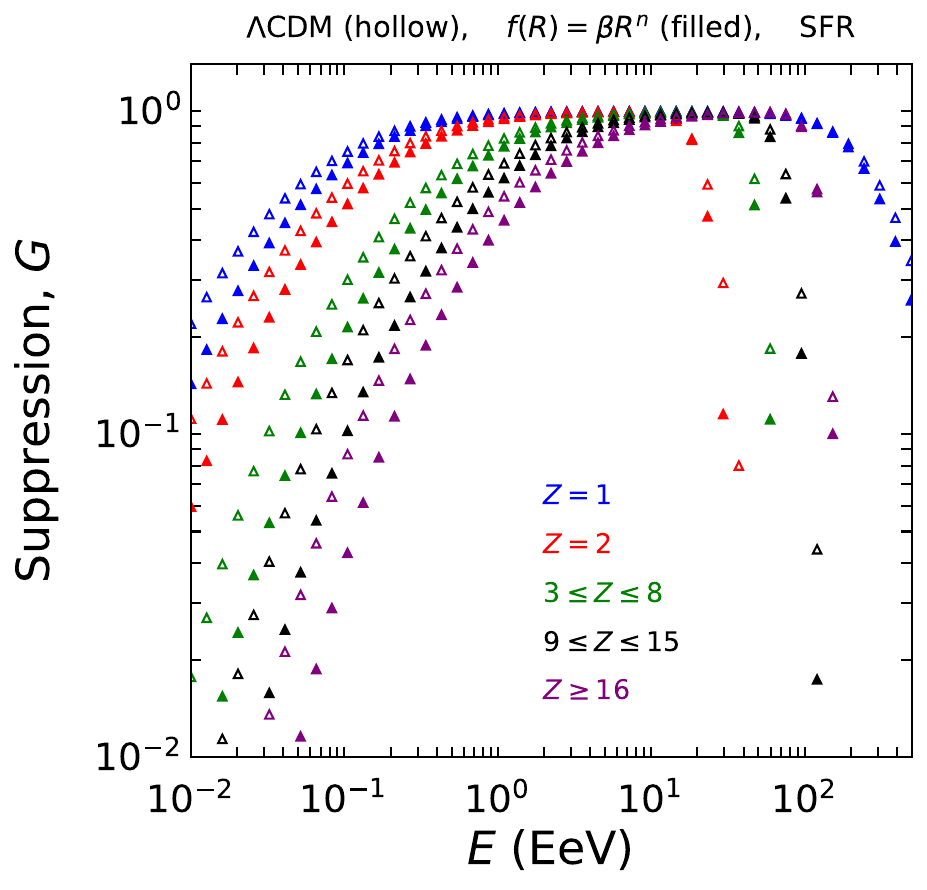}
}
\vspace{-0.3cm}
\caption{Left: Variations of suppression factor with respect to the energy 
$E$ for the different primaries as predicted by the $\Lambda$CDM model and 
the $f(R)$ gravity model with NE and $X_\text{s}=1$. Right: Same as the left 
pane but for the SFR case.}
\label{fig3}
\end{figure}

\begin{table}[htb!]
\caption{Parametrizations of Eq.~\eqref{fit} for the $\Lambda$CDM model and 
the $f(R)$ gravity power-law model with NE and SFR scenarios.}
    \vspace{5pt}
    \centering
    \begin{tabular}{c @{\hspace{0.5cm}} c @{\hspace{0.5cm}} c @{\hspace{0.5cm}} c @{\hspace{0.5cm}} c @{\hspace{0.5cm}} | @{\hspace{0.5cm}} c @{\hspace{0.5cm}} c @{\hspace{0.5cm}} c @{\hspace{0.5cm}} c @{\hspace{0.5cm}} c}
     \hline
     \hline
     \multicolumn{8}{c}{\hspace{0.7cm}NE} \\
     \hline
     \hline
     &&$\Lambda$CDM  &&&&$f(R)$ model \\
     \hline
     & a & b & $\vartheta$ & $\eta$ & a & b & $\vartheta$ & $\eta$ & \\
     \hline
     &0.224 & 0.162 & 1.541 & 0.140  &0.305 & 0.262 & 1.55 & 0.210 &\\
     \hline 
     \hline
     \multicolumn{8}{c}{\hspace{0.7cm}SFR} \\
     \hline
     \hline
     &&$\Lambda$CDM  &&&&$f(R)$ model \\
     \hline
     & a & b & $\vartheta$ & $\eta$ & a & b & $\vartheta$ & $\eta$ & \\
     \hline
     &0.215 & 0.334 & 1.810 & 0.270  &0.202 & 0.163 & 1.800 & 0.190 & \\
     \hline
    \end{tabular}
    \label{table1}
\end{table}
As we can see from Figs.~\ref{fig1} and \ref{fig2}  a change in suppression 
at higher energies is also evident and this suppression is mainly due to the 
radiation backgrounds. When the energy exceeds $60$ EeV, the attenuation 
length of protons decreases rapidly due to the photopion production with the 
CMB, falling significantly below $100$ Mpc for energies greater than $100$ 
EeV \cite{Manuel}. A similar phenomenon occurs for heavier nuclei at energies 
exceeding approximately $5Z$ EeV, as a result of the photodisintegrations with 
the CMB photons \cite{Manuel}. In simple words, when the energy reaches a point 
where the attenuation length is about the same or less than the distance to 
the nearest sources, we see a significant decrease in the flux. This scenario 
is seen clearly in Fig.~\ref{fig3}. For this computation, we take the 
different primaries from $Z=1$ upto $Z=26$. We consider $Z=7$ for the CNO 
group, $Z=14$ for the Si group, and $Z=26$ for the Fe group. The left panel of 
Fig.~\ref{fig3} is for the NE scenario, while the right panel is for the SFR 
by considering $X_\text{s}=1$. It is to be noted that the cosmological model 
effect is more evident in SFR than in the NE case. Here also the same results 
are obtained as in the earlier case that the $f(R)$ model exhibits 
a lower suppression in both high and low energy regimes than the $\Lambda$CDM 
model.
 
\begin{figure}[htb!]
\centerline{
\includegraphics[scale=0.42]{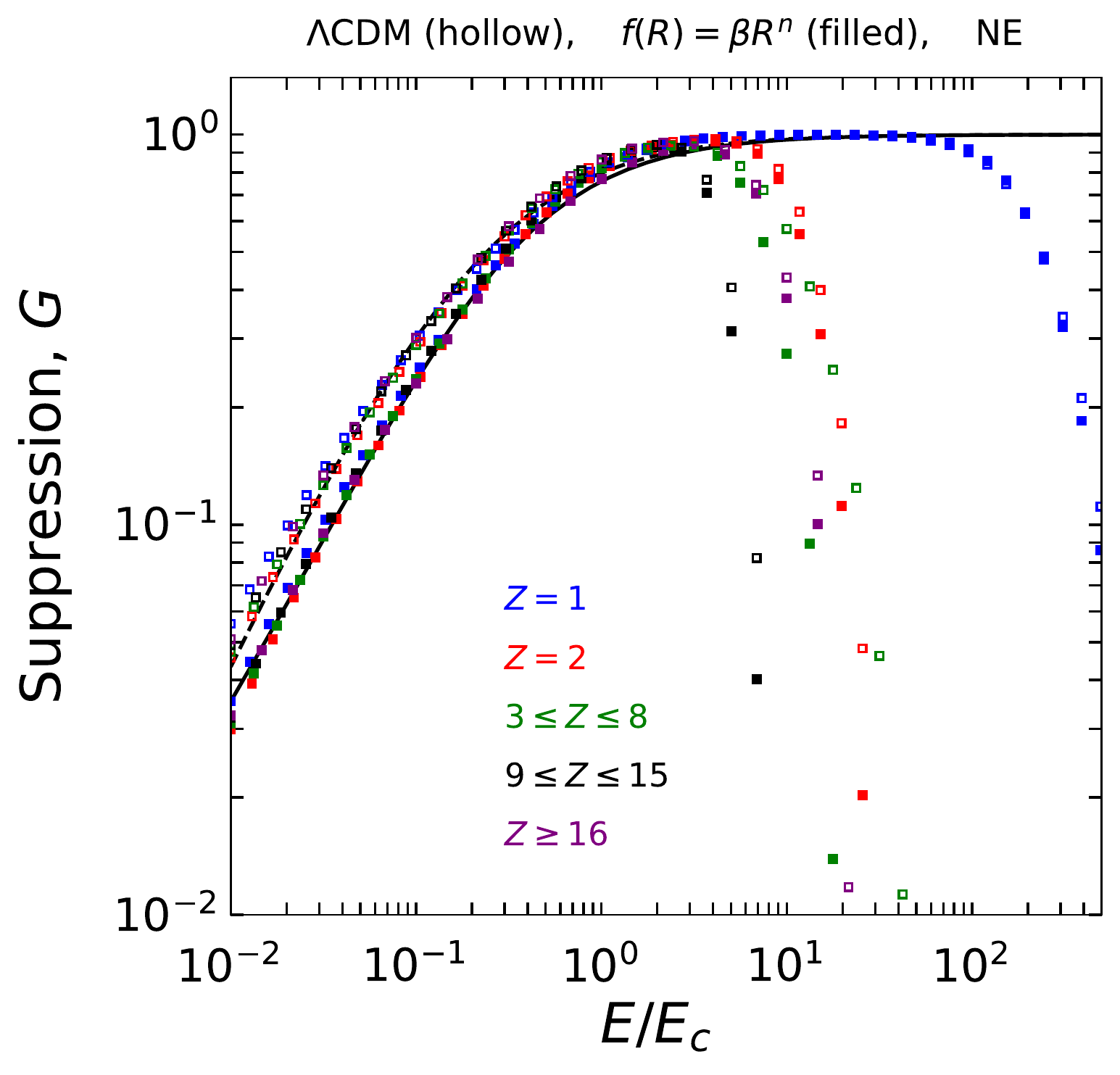}\hspace{1cm} 
\includegraphics[scale=0.42]{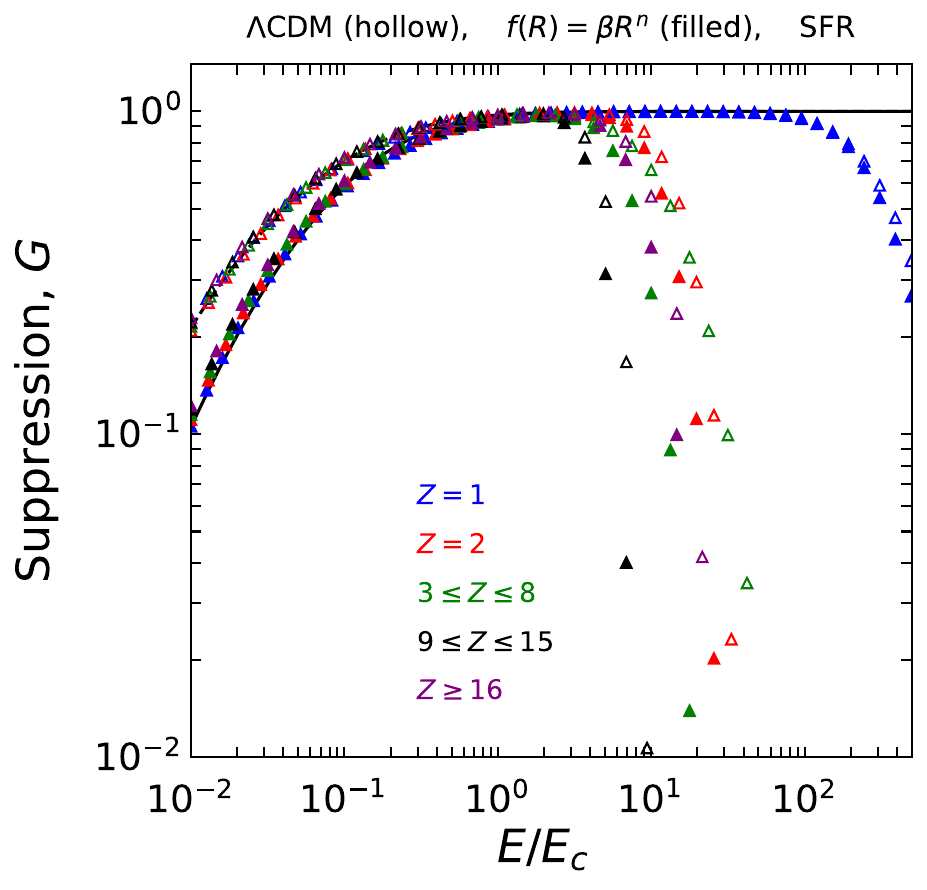}
}
\vspace{-0.3cm}
\caption{Left: Variations of suppression factor with respect to 
$E/E_\text{c}$ for the different primaries as predicted by the $\Lambda$CDM 
model and the $f(R)$ gravity model with NE and $X_\text{s}=1$. Right: Same as 
the left pane but for the SFR case.}
\label{fig4}
\end{figure}

In Fig.~\ref{fig4}, we also plot the suppression factor with respect to 
$E/E_\text{c}$ for different nuclei. In the left panel, we plot the NE 
case, while the SFR case is plotted in the right panel. 
%We divide the mass number by the critical energy $E_\text{c}$. 
We can see that all nuclei show the same behaviour for their respective 
primaries. Again here the nuclei in the $f(R)$ gravity model are less 
suppressed in comparison with the $\Lambda$CDM model. In the higher energy 
range, a sharp drop in flux appears which is mainly due to the 
photodisintegration. 
%\iffalse
%Another scenario is also possible that all nuclei exhibit their secondary 
%nuclei along with the primary ones. This scene is shown in Fig.~\ref{fig5}. 
%The empty squares are for the $\Lambda$CDM model and the filled squares are 
%for the $f(R)$ gravity model. Here we consider four mass groups 
%(He, N, Si, and Fe) and for each group, we plot for its corresponding 
%\fi

\begin{figure}[htb!]
\centerline{
\includegraphics[scale=0.42]{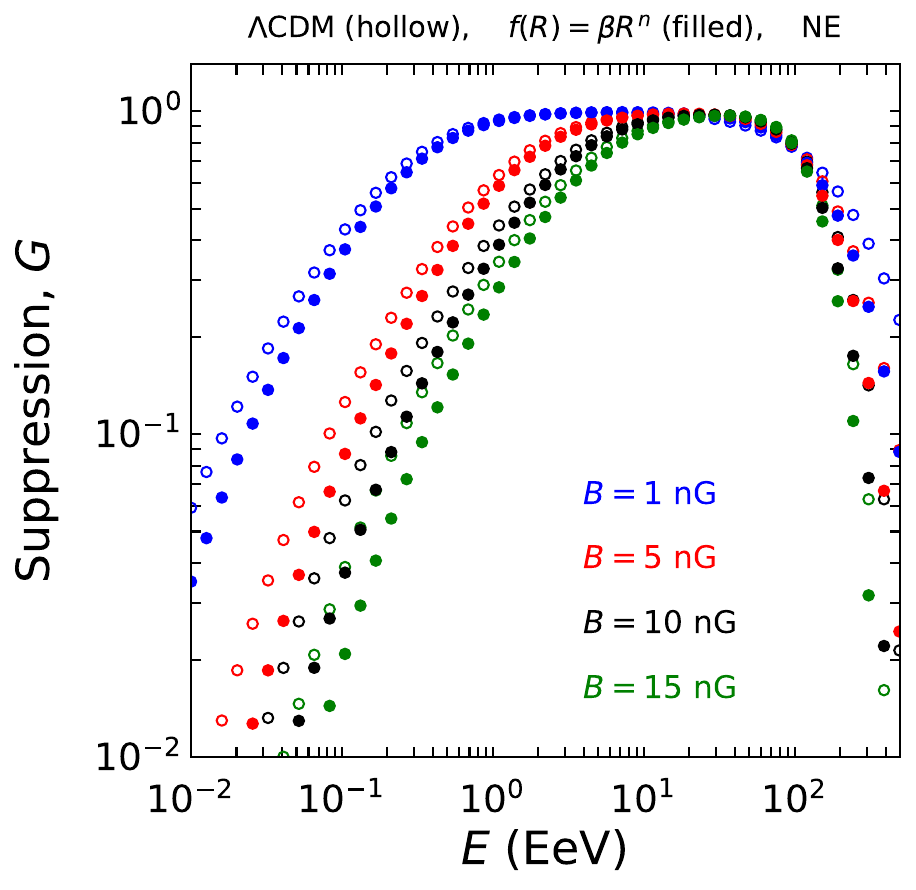}\hspace{1cm} 
\includegraphics[scale=0.42]{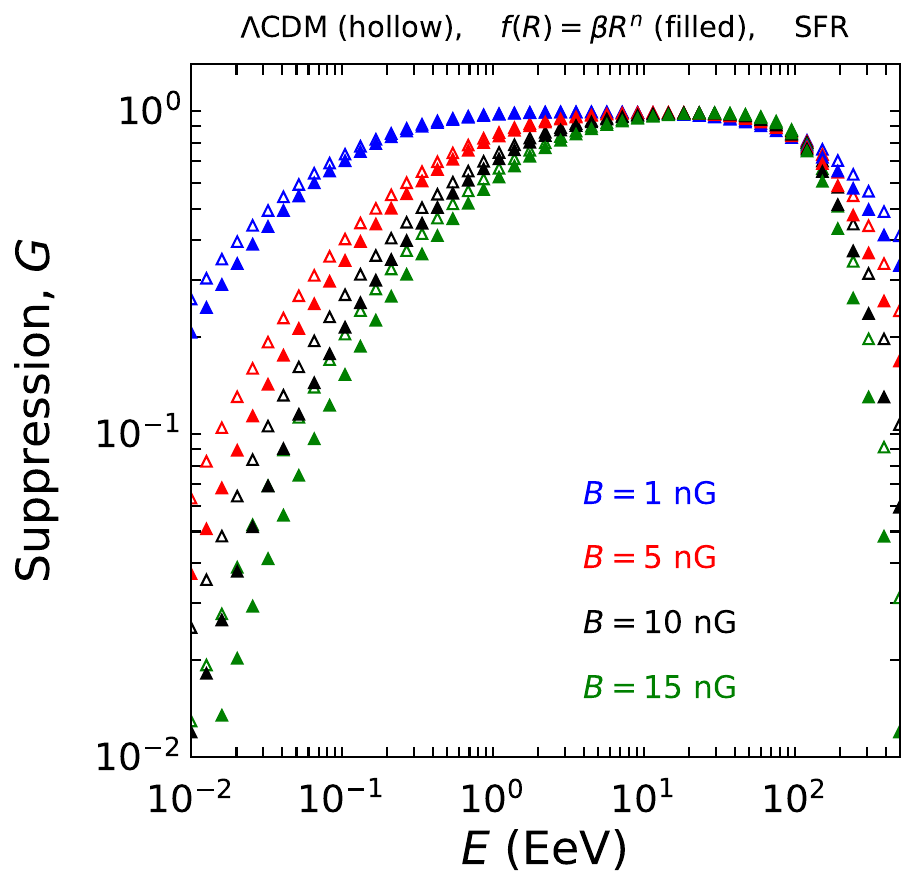}
}
\vspace{-0.3cm}
\caption{Left: Variations of suppression factor with respect to energy $E$ for 
different strengths of the magnetic field as predicted by the $\Lambda$CDM 
model and the $f(R)$ gravity power-law model with NE and $X_\text{s}=1$. 
Right: Same as the left pane but for the SFR case.}
\label{fig7}
\end{figure}

In all the above scenarios, we consider the magnetic field strength $B=1$ nG. 
Now, the effect of a variation of the magnetic field strength in the 
suppression factor can be seen in Fig.~\ref{fig7}. Here we consider the 
magnetic field strength from $1$ nG to $15$ nG. In this figure, the left panel 
is for the NE case while the right one is for the SFR case. The cosmological 
model effect is the same as that of the earlier cases. At low energy levels, 
the predictions of the  $\Lambda$CDM  model at $15$ nG closely align with 
those of the $f(R)$ gravity power-law model at $10$ nG. The variation of 
suppression factor between magnetic fields of strengths $1$ nG to $5$ nG is 
quite large as compared to that for the $5$ nG to $15$ nG. We also check (not 
included here) for the higher magnetic field strength such as $50$ nG, but 
interestingly no significant variation can be seen as we see the variation 
for the $1$ nG to $5$ nG.

\begin{figure}[htb!]
\centerline{
\includegraphics[scale=0.42]{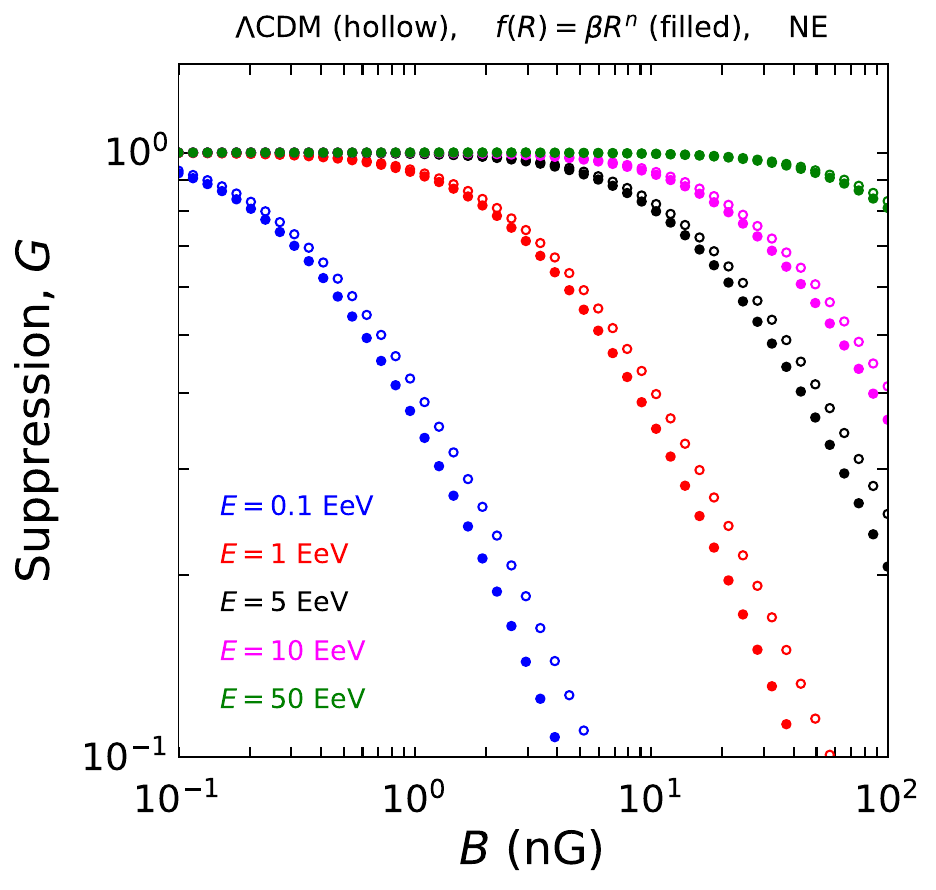}\hspace{1cm} 
\includegraphics[scale=0.42]{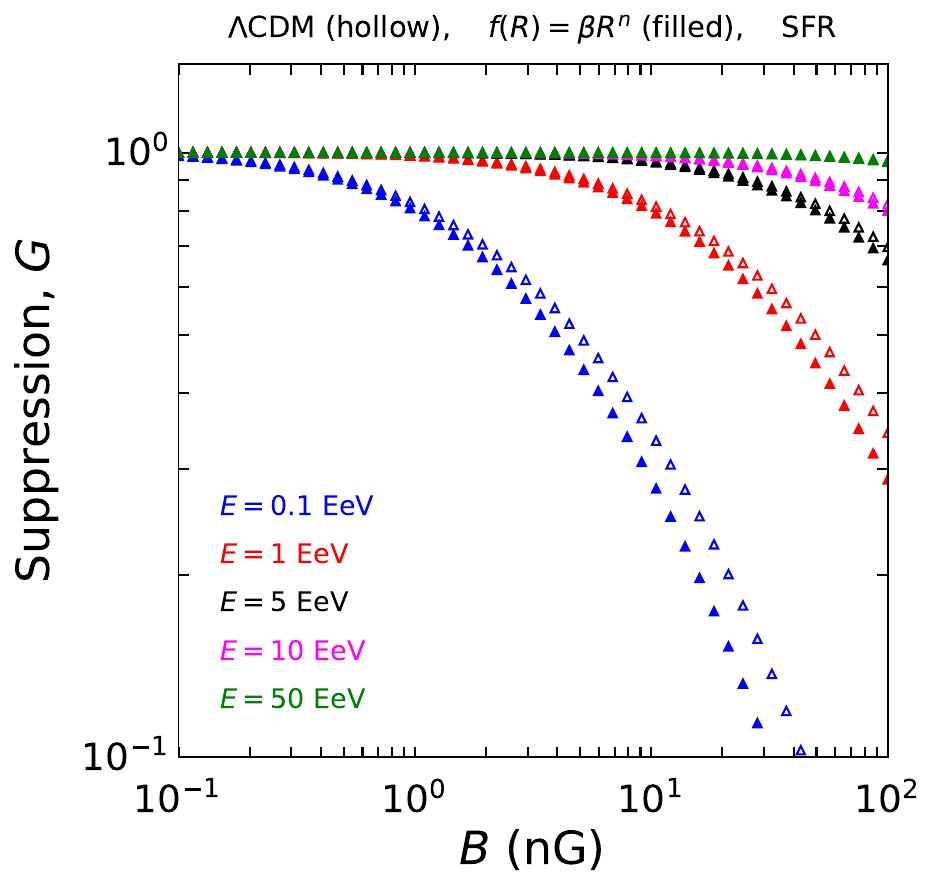}
}
\vspace{-0.3cm}
\caption{Left: Variations of suppression factor with respect to the magnetic 
field for different energy values as predicted by the $\Lambda$CDM and the 
$f(R)$ gravity power-law model with NE and $X_\text{s}=1$. Right: Same as the 
left pane but for the SFR case.}
\label{fig8}
\end{figure}
For a more clear picture of the variation of the suppression factor with 
magnetic field strength, in Fig.~\ref{fig8}, the flux suppression factor as a 
function of the magnetic field is plotted. For this scene, we take the energy 
values from $E=0.1$ EeV to $E=50$ EeV, and the NE and SFR cases are also shown 
in the left and right panels of this figure respectively. The standard 
$\Lambda$CDM model predicts higher suppression than the $f(R)$ gravity 
power-law model as expected. The cosmological model's effect is evident 
throughout the plot, but as the TMF shifts to a higher strength, the model 
effect becomes more. It is seen that the suppression decreases quickly with 
increasing magnetic field for low energy particles and for substantially high 
energy particles it almost remains constant up to very high magnetic field 
strength. One important result to be noted is that the suppression predicted 
by the $f(R)$ power-law model at $E=1$ EeV in the NE case is the same as the 
$\Lambda$CDM at $E=0.1$ EeV in the SFR case.   

\subsection{Results from f(Q) gravity model}\label{secVB}
\begin{figure}[htb!]
\centerline{
\includegraphics[scale=0.42]{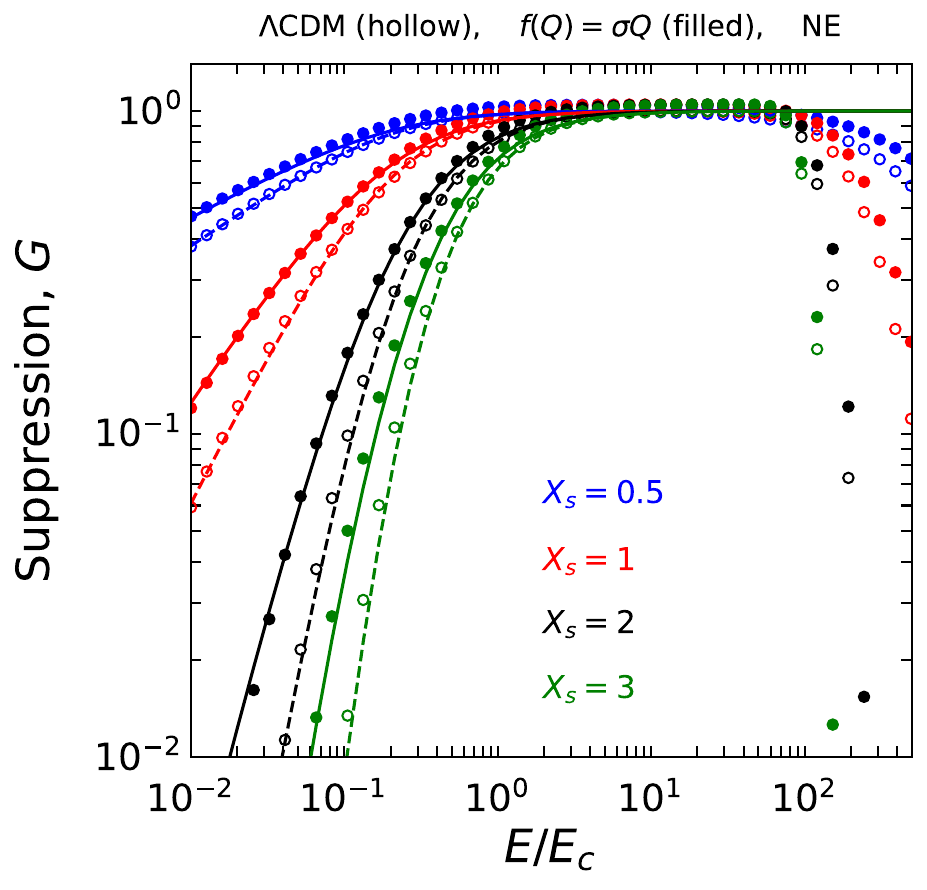}\hspace{1cm} 
\includegraphics[scale=0.42]{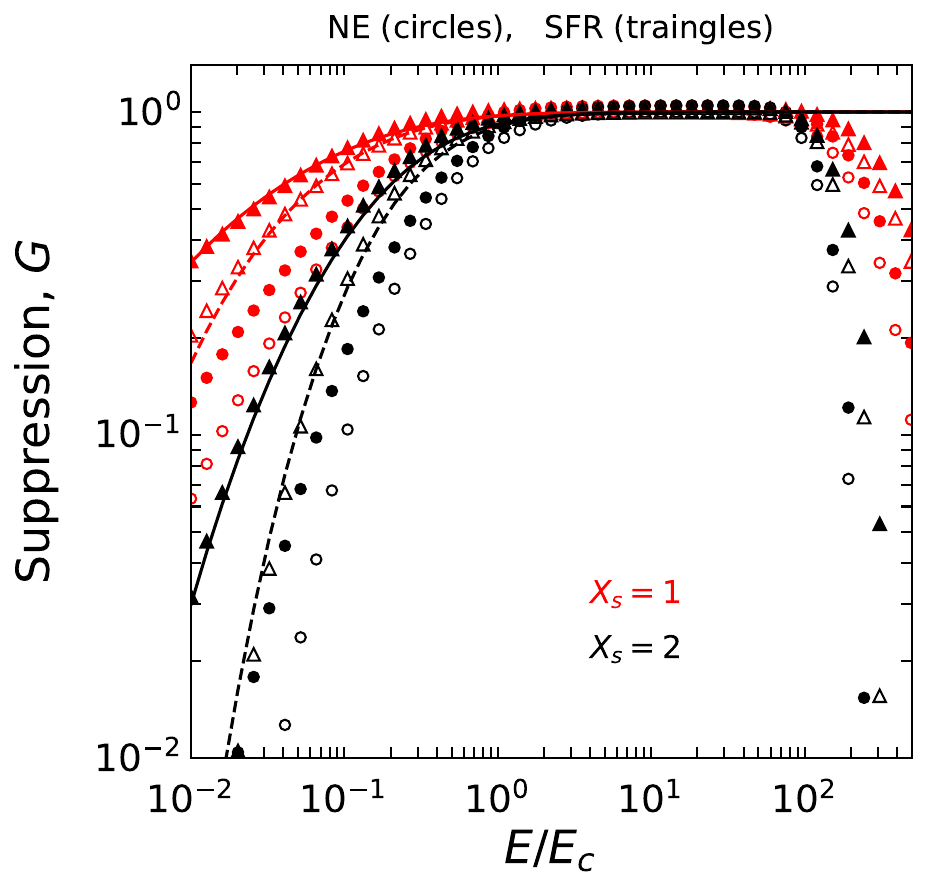}
}
\vspace{-0.3cm}
\caption{Left: Suppression factors with respect to $E/E_\text{c}$ for 
different density factors of sources $X_\text{s}$ as predicted by the 
$\Lambda$CDM model and the $f(R)$ gravity model. Right: The suppression 
factors for two different density factors of sources $X_\text{s}=1$ and 
$X_\text{s}=2$ as predicted by the $\Lambda$CDM 
model and the $f(R)$ gravity model. The round and triangle-shaped points 
represent the NE and SFR scenarios respectively. The hollow and filled points 
represent the results from the $\Lambda$CDM model and the $f(Q)$ gravity model 
respectively. The dotted and solid lines are for the analytic fitting for 
these two respective cosmological models.}
\label{fig9}
\end{figure}

Following the same procedures as discussed above, here we discuss the 
numerical results obtained from the $f(Q)$ gravity model. Fig.~\ref{fig9} 
shows the variation of the suppression factor with respect to $E/E_\text{c}$ 
as predicted by the $f(Q)$ gravity model in comparison with that of the
$\Lambda$CDM model. In the left panel of the figure, we have shown the NE 
scenario while the SFR is shown in the right panel. In the plots, the empty 
circles or triangles are for the $\Lambda$CDM model while the filled ones are 
for the $f(Q)$ gravity model.

\begin{table}[htb!]
\caption{Parametrizations of Eq.~\eqref{fit} for the $f(Q)$ gravity model 
with NE and SFR scenario, and with $\gamma = 2$.}
    \vspace{5pt}
    \centering
    \begin{tabular}{c @{\hspace{1.5cm}} c @{\hspace{1.5cm}} c @{\hspace{1.5cm}} c @{\hspace{1.5cm}} c @{\hspace{1.5cm}} c}
     \hline
     \hline
     \multicolumn{5}{c}{NE} \\
     \hline
     & a & b & $\vartheta$ & $\eta$ & \\
     \hline
     &0.194 & 0.172 & 1.448 & 0.160  &\\
     \hline 
     \hline
     \multicolumn{5}{c}{SFR} \\
     \hline
     & a & b & $\vartheta$ & $\eta$ & \\
     \hline
     &0.175 & 0.301 & 1.710 & 0.231 &\\
     \hline
    \end{tabular}
    \label{table3}
\end{table}
Unlike the previous case of MTG, here the $f(Q)$ gravity model depicts a 
higher suppression than that of the $\Lambda$CDM model. In the right panel, 
the red points are for the density factor of sources $X_\text{s}=1$ and the 
black points are for $X_\text{s}=2$. The solid and dotted lines are for the 
corresponding fitting of the numerical results of the $f(Q)$ gravity model 
and the $\Lambda$CDM model. We adopt the same Eq.~\eqref{fit} for the analytic 
fitting. The cosmological model's effect is more evident in the SFR case as 
compared to the NE case. The fitting parameters relevant to this calculation 
are given in Table \ref{table3}.

\begin{figure}[htb!]
\centerline{
\includegraphics[scale=0.42]{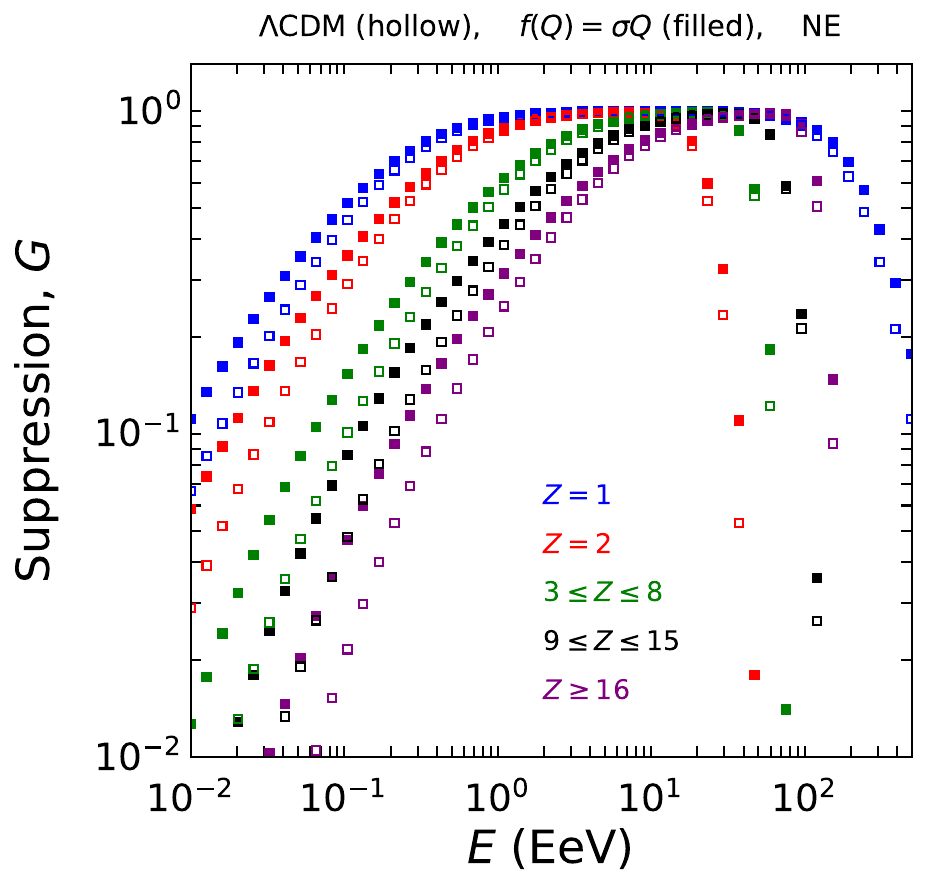}\hspace{1cm} 
\includegraphics[scale=0.42]{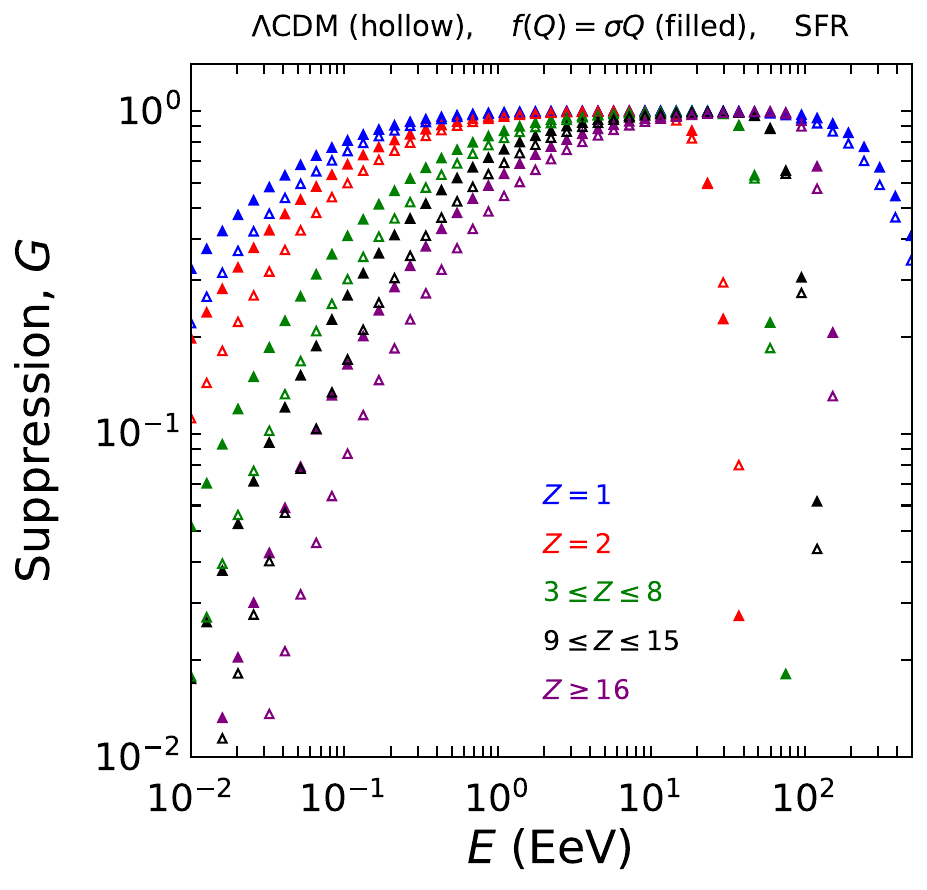}
}\vspace{0.5cm}
\centerline{
\includegraphics[scale=0.42]{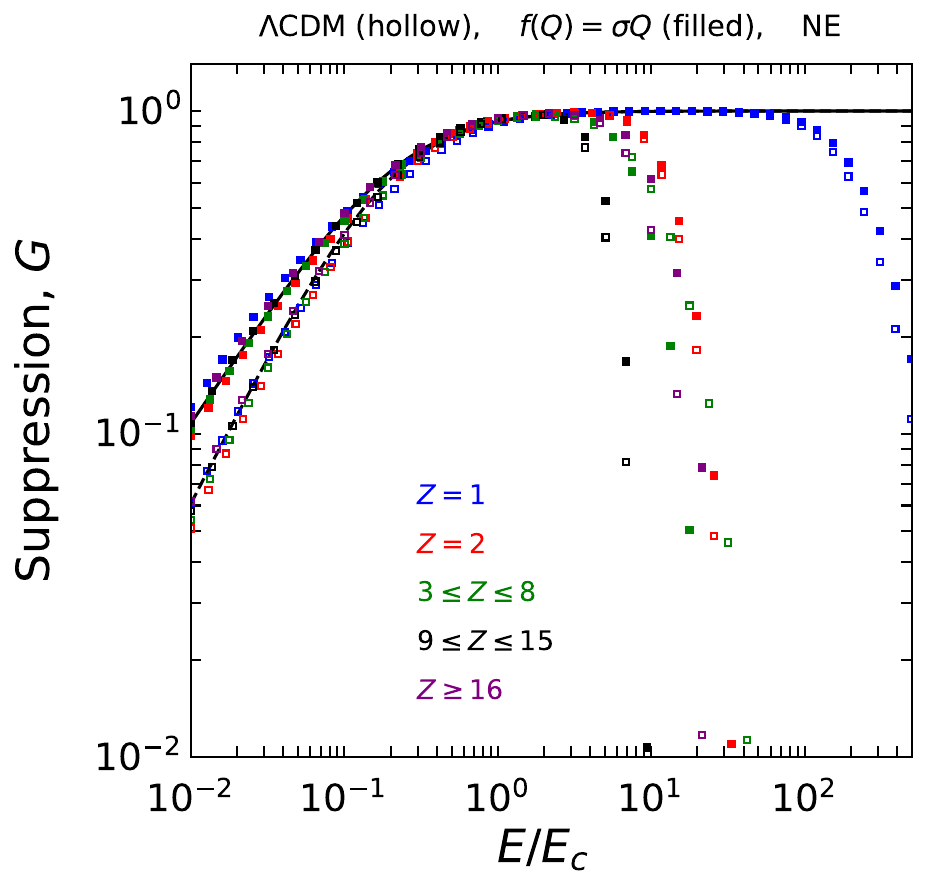}\hspace{1cm} 
\includegraphics[scale=0.42]{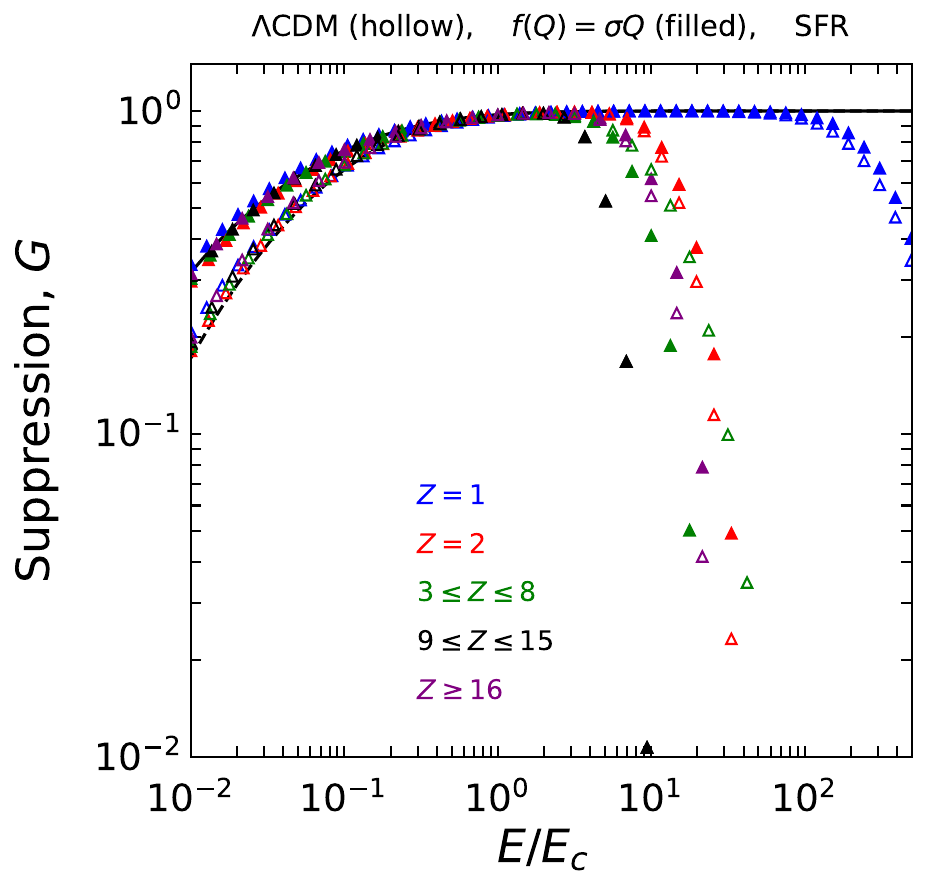}
}
\vspace{-0.3cm}
\caption{Variations of suppression factor $G$ with respect to the energy $E$ 
(top panels) and $E/E_\text{c}$ (bottom panels) for the different nuclei as
predicted by the $\Lambda$CDM model and the $f(Q)$ gravity model with NE and 
$X_\text{s}=1$.}
\label{fig10}
\end{figure}

A mixed composition of nuclei upto iron is plotted in Fig.~\ref{fig10}. The 
left panels of Fig.~\ref{fig10} are for the NE scenario, while the right 
panels are for the SFR by considering $X_\text{s}=1$. Here also the same 
results are obtained as in the earlier case that the $f(Q)$ model exhibits a 
higher suppression in both high and low energy regimes than the $\Lambda$CDM 
model. In the top panels of Fig.~\ref{fig10}, we plot the suppression factor 
with respect to the energy for different nuclei. In the bottom panels of 
Fig.~\ref{fig10}, we also plot the suppression factor with respect to 
$E/E_\text{c}$ for the same nuclei. In the left panel, we plot the NE 
case, while the SFR case is plotted in the right panel. We can see that all 
nuclei show the same behaviour for their respective primaries when plotting 
with respect to $E/E_\text{c}$. Again here the nuclei in the $\Lambda$CDM model 
are less suppressed in comparison with the $f(Q)$ model. A sharp drop in flux 
appears in higher energy cases due to the photodisintegration as we have 
already discussed in the previous section. 

\begin{figure}[htb!]
\centerline{
\includegraphics[scale=0.42]{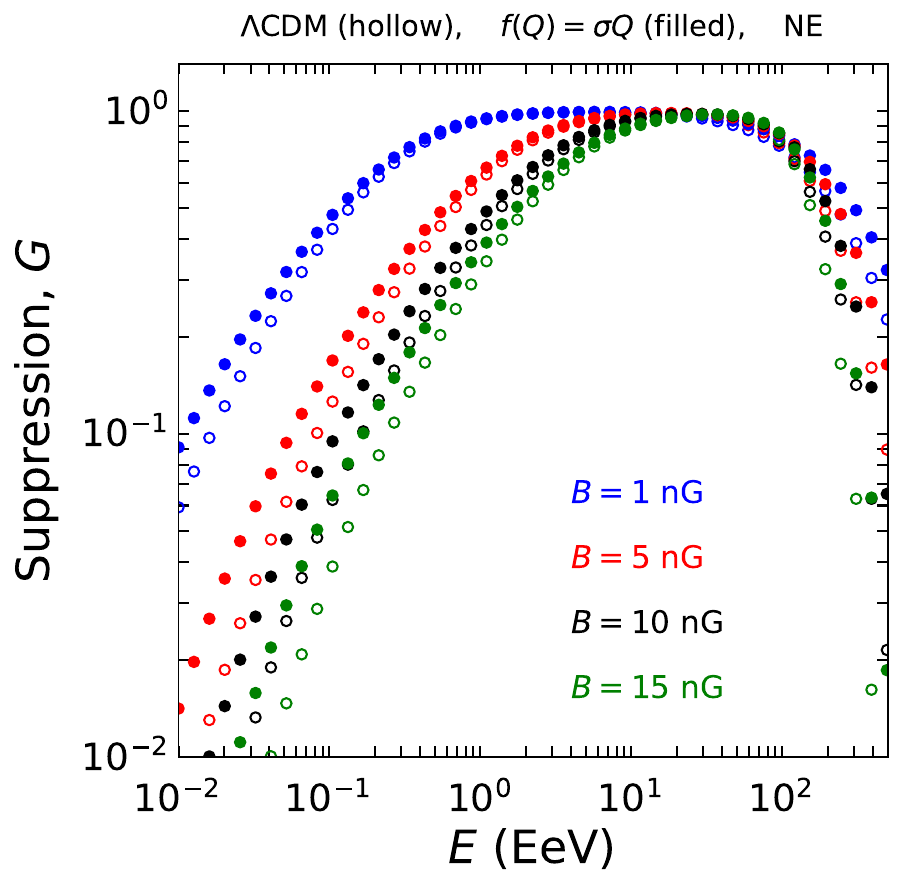}\hspace{1cm} 
\includegraphics[scale=0.42]{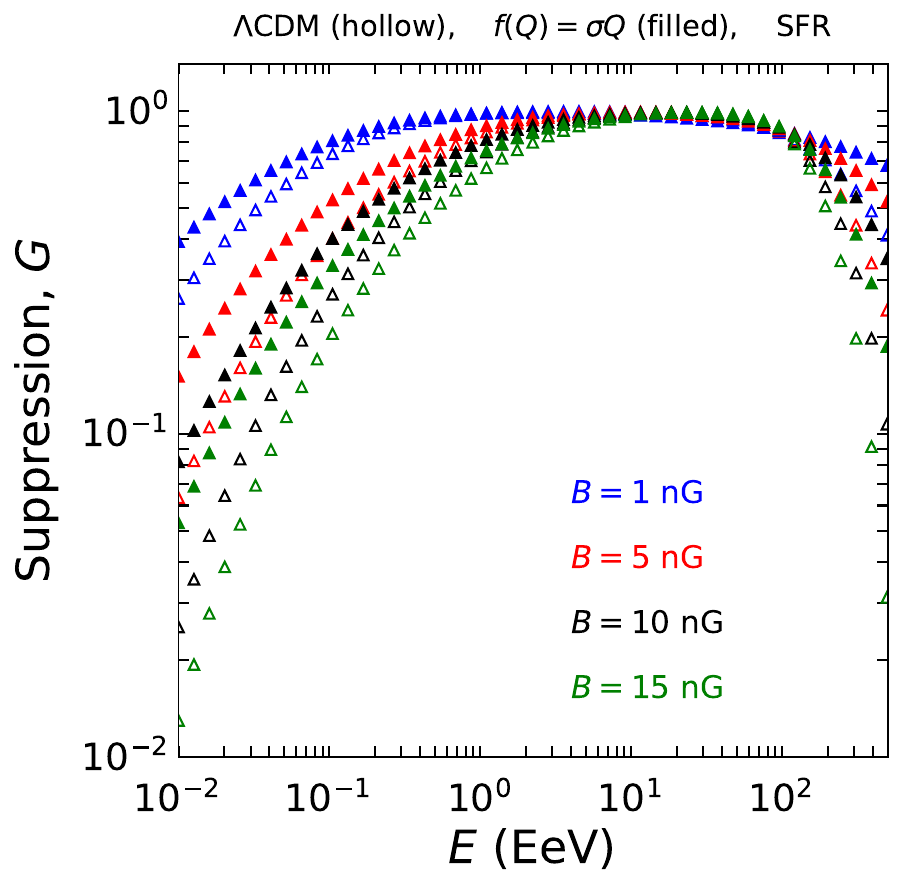}
}
\vspace{-0.3cm}
\caption{Left: Variations of suppression factor with respect to the energy $E$ 
for the different strengths of the magnetic field as predicted by the 
$\Lambda$CDM and the $f(Q)$ gravity model with NE and $X_\text{s}=1$. 
Right: Same as the left pane but for the SFR case.}
\label{fig12}
\end{figure}
\begin{figure}[htb!]
\centerline{
\includegraphics[scale=0.42]{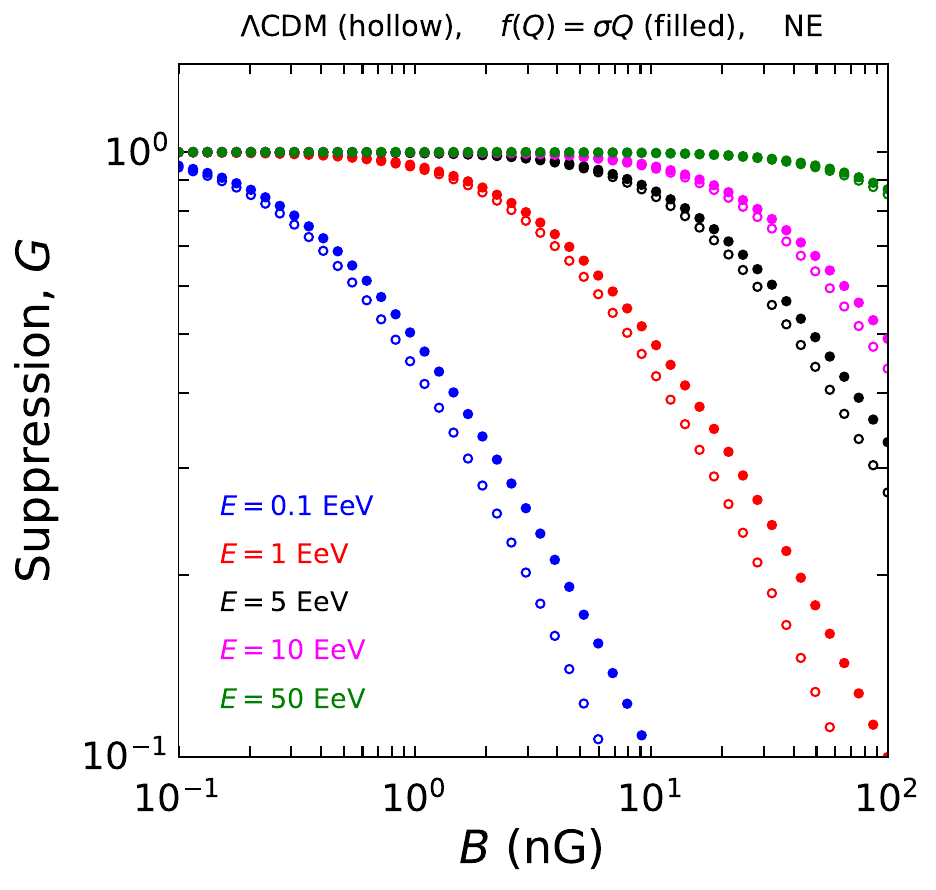}\hspace{1cm} 
\includegraphics[scale=0.42]{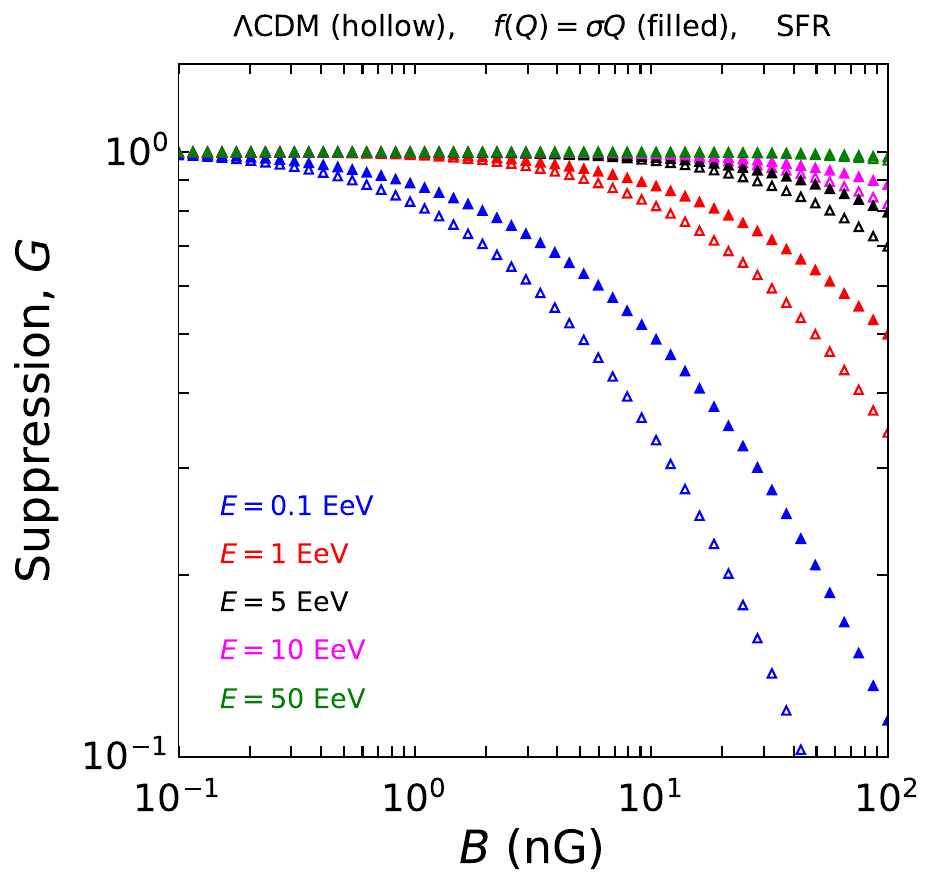}
}
\vspace{-0.3cm}
\caption{Left: Variations of suppression factor with respect to the magnetic field for the different energy values as predicted by the $\Lambda$CDM and 
the $f(Q)$ gravity model with NE and $X_\text{s}=1$. Right: Same as the left 
pane but for the SFR case.}
\label{fig13}
\end{figure}

A discernible variation in the suppression factor with respect to the magnetic 
field strength is illustrated in Fig.~\ref{fig12}. As in the previous case, 
the magnetic field strength under consideration ranges from $1$ nG to $15$ nG. 
In Fig.~\ref{fig12}, the left panel corresponds to the NE case, while the 
right panel represents the SFR case. The influence of the cosmological model 
is consistent with previous cases. At lower energy levels, the $\Lambda$CDM 
model's predictions at $10$ nG align closely with those of the $f(Q)$ model 
at $15$ nG. We will discuss these types of scenarios in the next section.  
The suppression factor exhibits a substantial variation between the magnetic 
field of strengths $1$ nG and $5$ nG, compared to the range from $5$ nG to 
$15$ nG. We also examined higher magnetic field strengths, such as $50$ nG 
(results not included here). Interestingly, the variation observed in this 
range was not as significant as the variation from $1$ nG to $5$ nG. In 
Fig.~\ref{fig13}, we present the flux suppression factor as a function of the 
strength of the magnetic field. For this analysis, we consider energy values 
ranging from $E=0.1$ EeV to $E=50$ EeV. The left and right panels of Fig.~\ref{fig13} depict the NE and SFR cases, respectively. The $f(Q)$ gravity 
model predicts a higher degree of suppression compared to the standard 
$\Lambda$CDM model. The influence of the cosmological model is apparent 
throughout the plot. However, as the TMF shifts to a higher strength, the 
model's effect becomes more. The CR flux suppression decreases quickly with 
increasing magnetic field strength for low-energy particles and it remains 
constant for substantially high-energy particles up to very high 
magnetic field strength as in the previous case.

\section{Conclusions}\label{secVI}
We study in detail the CR flux suppression in both low and high-energy 
regions through the magnetic horizon effect with the diffusion effect of CRs 
within the TMF. We consider a model for each of two MTGs for this analysis viz.\ the 
$f(R)$ gravity and the $f(Q)$ gravity. In the left panel of Fig.~\ref{fig0}, 
we have shown the difference of the cosmological models through the Hubble 
parameter expression and in the right panel the cosmological time evolution 
with the redshift. We have calculated the magnetic suppression factor for all 
the cosmological models including the $\Lambda$CDM by considering the NE and 
SFR scenarios for different density sources $X_\text{s}$ in Fig.~\ref{fig1} 
and \ref{fig2}. A scenario of the nuclei flux suppression as a function 
of both $E$ and $E/E_\text{c}$ is presented in detail in Fig.~\ref{fig3} and 
Fig.~\ref{fig4}. We parameterize all these numerical results through 
an analytical factor $P(E/E_\text{c})$ given in Eq.~\eqref{fit}. We consider 
the effect of the magnetic field in the suppression factor for all the 
cosmological models in Figs.~\ref{fig7} and \ref{fig8}. We discuss all of the  
above cases for the $f(Q)$ gravity model in terms of Figs.~\ref{fig9} -- 
\ref{fig13} including NE and SFR scenarios. The effect of the cosmological 
model is better distinguished in the SFR case as compared to the case of NE.

\begin{figure}[htb!]
\centerline{
\includegraphics[scale=0.42]{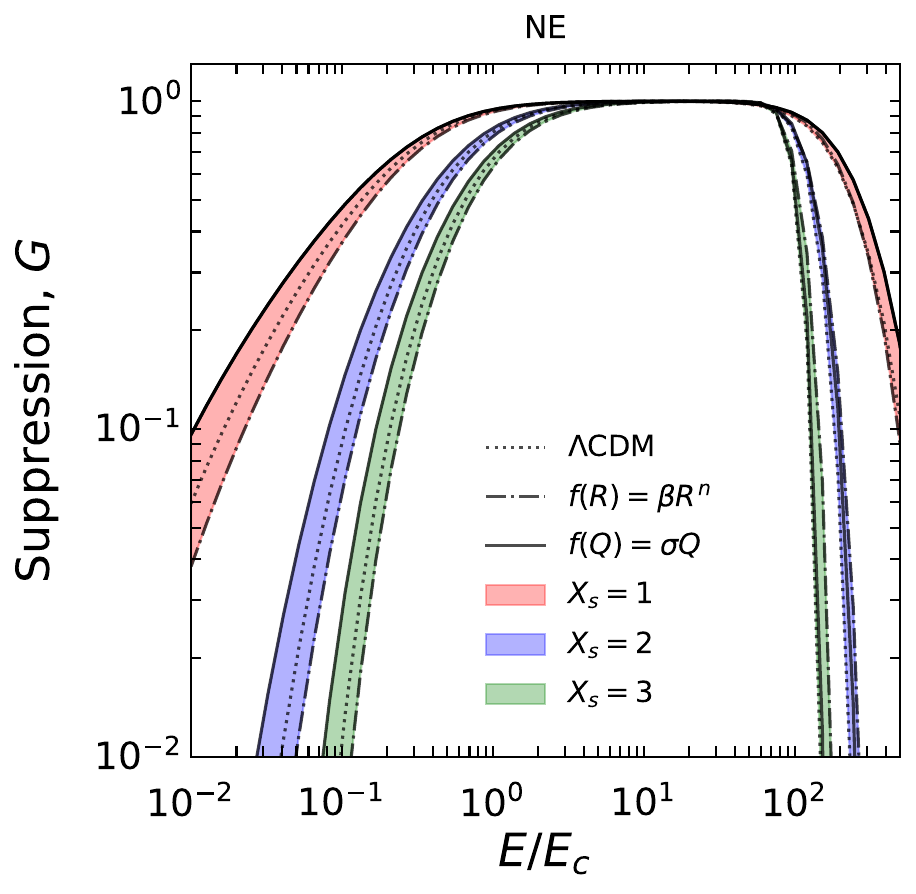}\hspace{1cm}
\includegraphics[scale=0.42]{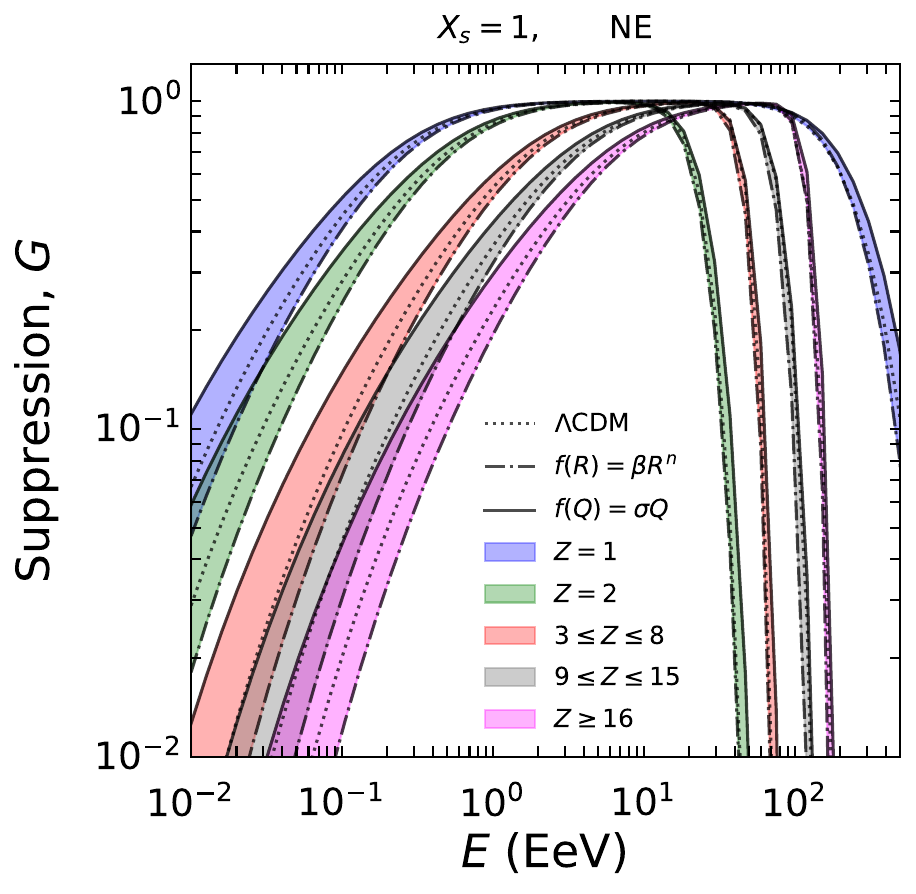}
}\vspace{0.5cm}
\centerline{
\includegraphics[scale=0.42]{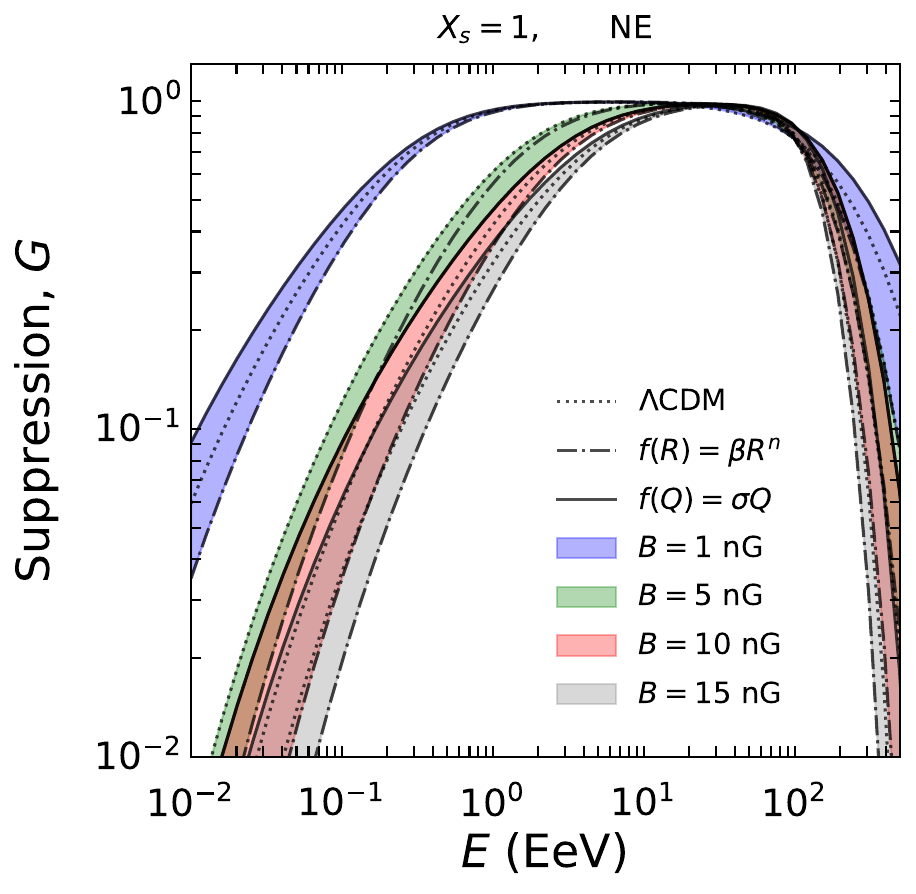}\hspace{1cm}
\includegraphics[scale=0.42]{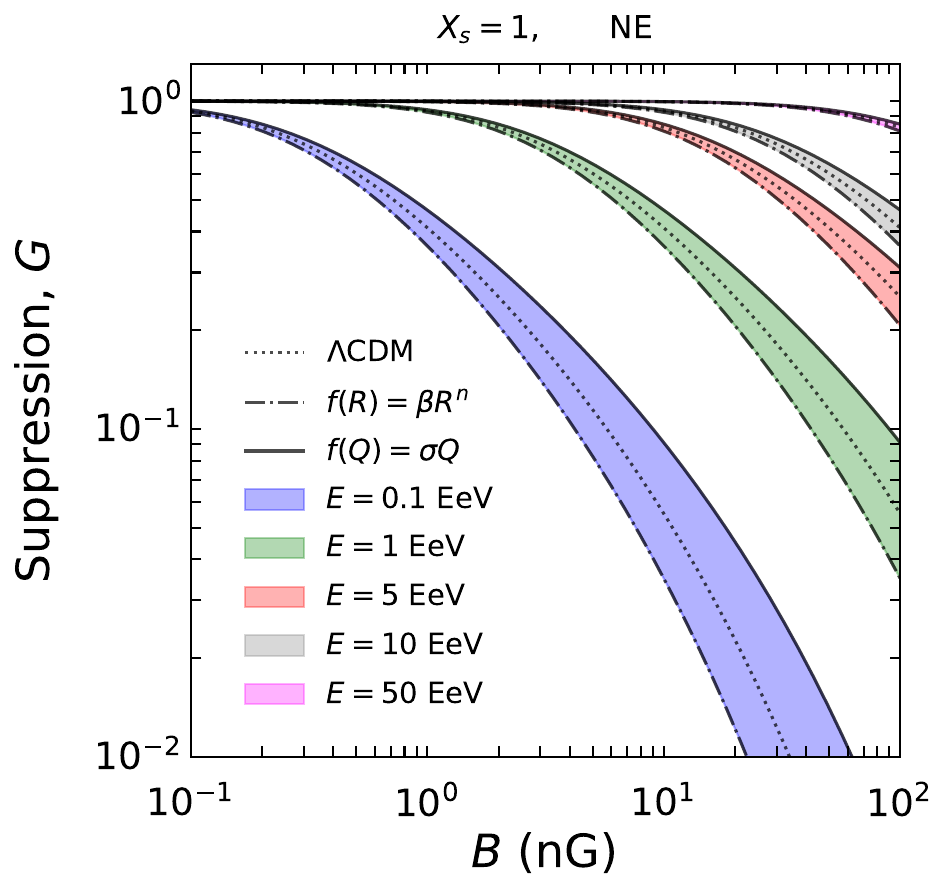}
}
\vspace{-0.3cm}
\caption{Suppression factors for the different scenarios (see text) for the 
$\Lambda$CDM (dotted lines), $f(R)$ gravity power-law (dash-dot lines), and 
$f(Q)$ gravity model (solid lines) in NE case.}
\label{figc}
\end{figure}

In Fig.~\ref{figc}, we summarize the results of all three cosmological models 
for the study of the suppression factor in the case of NE (all panels) and 
$X_\text{s}=1$ (top right, bottom left and right panels). Since we have 
observed the same kind of patterns in SFR cases also, we have not included 
this scenario here. In the shaded regions, the upper boundary line denotes 
the results from the $f(Q)$ gravity 
model (solid line), while the lower boundary line denotes the $f(R)$ model 
(dash-dot line). The standard $\Lambda$CDM model (dotted line) lies between 
these two. From the top left panel, we can see that as the finite density of 
sources $X_\text{s}$ increases, the cosmological model's effect in the high 
energy regime becomes narrower. In the top right panel, a scenario of the 
suppression factor is shown for the different primary particles as predicted by
the considered cosmological models.
From the bottom left panel, we can see that in between the magnetic field 
strength of $10$ nG and $15$ nG, all three cosmological models exhibit a 
similar kind of results in both low and high-energy regimes. Again from the 
bottom right panel, one can see that at the high energy regime, all considered 
cosmological models behave the same. Thus from this Fig.~\ref{figc}, we can 
conclude that the $f(R)$ gravity power-law model exhibits low suppression 
results, the standard $\Lambda$CDM model exhibits a moderate suppression, and 
the $f(Q)$ gravity model performs a higher suppression results in the UHECR 
flux.

The observations from the Pierre Auger Observatory indicate that as energy 
increases, heavier elements become more dominant in the CRs composition. To 
avoid the overlapping between different elements, strong suppression of heavy 
elements is necessary at lower energies. While elementary spectra with a very 
hard spectral index at the sources ($\gamma < 1$) could explain this, it 
contradicts the expectations from second-order Fermi acceleration 
\cite{pao_jcap04, Manuel}. An alternative explanation discussed in this 
present study involves a magnetic suppression effect, leading to a hardened 
spectrum for low rigidities. This effect could account for the observed 
composition and spectrum patterns with the spectral index at 
$\gamma=2$ \cite{mollerach2013, mollerach_prd_2020}. The MTGs play an 
effective role in explaining the behaviours of CR flux suppression in the
Universe with accelerated expansion. The primary objective of this work is 
to investigate the potential impact of MTGs on the suppression of CR flux.
The various predictions put forth by these models highlight the need for 
further refinement in our comprehension of UHECR sources, their propagation, 
and the underlying principles of gravity. Moving forward, this study can be 
expanded by incorporating additional MTGs into both quantitative and more 
realistic analyses including secondary particles along with the primaries, 
aiming to gain deeper insights into the characteristics of CRs.

The extragalactic UHECRs that arrive from discrete sources differ from a 
continuous distribution across the space. At higher energies, the attenuation 
length of particles owing to interactions with the radiation backgrounds is 
comparable to the separation distance between the sources, resulting in a 
stronger suppression of flux. The background interactions can dramatically 
reduce the UHECR flux from nearby sources. This can have a significant impact 
on the CR flux recovery if the source spectrum extends beyond the attenuation 
limit \cite{Manuel}. Due to the magnetic horizon effect, the low energy flux 
gets suppressed from a discrete source distribution in the presence of a 
magnetic field. When particles like protons and nuclei come from space, they 
can be affected by the magnetic fields. If these particles have the same 
rigidity, which is related to their charge and energy, they will follow the 
same trajectory through these magnetic fields. Particles that keep 
their original masses experience more magnetic suppression when taken as 
a function of $E/E_c$, similar to what happens to protons \cite{Manuel}.
Interestingly, even when particles lose some of their mass and charge during 
their journey as a result of photodisintegration, they mostly keep the same 
rigidity. This is because as their mass and charge decrease together, keeping 
their ratio the same. So, these secondary particles follow a trajectory 
through the magnetic field that is similar to that of the primary particles 
that exhibit without photodisintegration. We will study these secondary 
particles generated for different cosmological models in our next work.

\section*{Acknowledgements} 
SPS is grateful to the authors of CRPropa and SimProp for making it available. 
UDG is thankful to the Inter-University Centre for Astronomy and Astrophysics 
(IUCAA), Pune, India for the Visiting Associateship of the institute.

\end{document}